\documentclass[1p,12pt]{elsarticle}

\usepackage{hyperref}
\usepackage{longtable}
\usepackage[dvipsnames]{xcolor}
\usepackage[export]{adjustbox}
\usepackage{bm}
\usepackage{xcolor}
\usepackage{amsmath}
\usepackage{cleveref} 
\usepackage{parskip}
\usepackage{amssymb}

\newcommand{\spac}{{\hspace{0.3mm}}}

\usepackage{physics}

\usepackage{lineno}

\begin{document}

\begin{frontmatter}

\title{A Time Projection Chamber to Search for Feebly Interacting Bosons via Proton Induced Nuclear Reactions}

\author[label1]{Martin Sevior}
\address[label1]{School of Physics, The University of Melbourne, Parkville, Victoria, 3010, Australia}
\author[label2]{Michael Baker}
\address[label2]{ARC Centre of Excellence for Dark Matter Particle Physics, School of Physics, The University of Melbourne, Victoria 3010, Australia}
\author[label3]{Lindsey Bignell}
\address[label3]{Department of Nuclear Physics and Accelerator Applications, The Australian National University, Canberra ACT, 2600, Australia}
\author[label4]{Catalina Curceanu} 
\address[label4]{Laboratori Nazionali di Frascati dell’INFN, Via E. Fermi 54 (gia’ 40), 00044 Frascati (Roma), Italy}
\author[label3]{Jackson T.H. Dowie}
\author[label3]{Tibor Kibédi} 
\author[label1]{David Jamieson} 
\author[label3]{Andrew Stuchbery} 
\author[label1]{Andrea Thamm} 
\author[label5]{and Martin White}
\address[label5]{School of Physics, Chemistry and Earth Sciences, University of Adelaide, South Australia
5005 Australia }
\begin{abstract}
 We propose a new Time Projection Chamber particle detector (TPC) to search for the existence of feebly-interacting bosons and to investigate the existence of the X17 boson, proposed by the ATOMKI group to explain anomalous results in the angular distributions of $e^{+}e^{-}$ pairs created in proton-induced nuclear reactions. Our design will provide 200 times greater sensitivity than ATOMKI and the program of research will also provide world-leading limits on feebly interacting bosons in the mass range of 5 - 25 MeV.    
\end{abstract}

\begin{keyword}
Particle Physics, New Physics, Time Projection Chamber, Nuclear Physics
\end{keyword}

\end{frontmatter}

\newpage
\section{Introduction}

There has recently been renewed interest in feebly interacing particles with sub-GeV masses, partly due to the null results of the LHC and partly due to new experimental opportunities. These particles are motivated by a wide variety of theoretical models that address open problems in the Standard Model (SM) such as the hierarchy problem, the strong CP problem, neutrino oscillations and the existence of dark matter (DM) (for a recent review see \cite{Lanfranchi:2020crw,Agrawal:2021dbo} and references theirin). Feebly interacting particles with masses between an MeV to several GeV are targeted by a variety of fixed-target, beam dump, collider, and accelerator-based neutrino experiments. Among these experiments is the ATOMKI collaboration which recently found evidence for a feebly interacting boson, the X17, with a mass of 17 MeV produced in three separate nuclear reaction experiments: 
\begin{itemize}
    \item $p\, + ^{7}$Li$ \, \to \,^{8}$Be$^* \,\to\, ^{8}$Be$ \,+ \rm{X17}(e^{+}e^{-})$\cite{2016Kr_PRL},
    \item $p\, + ^{3}$H$ \, \to \,^{4}$He$^* \,\to\, ^{4}$He$ \,+ \rm{X17}(e^{+}e^{-})$\cite{2021Kr02_PRC}, and
    \item $p\, + ^{11}$B$ \, \to \,^{12}$C$^* \,\to\, ^{12}$C$ \,+ \rm{X17}(e^{+}e^{-})$\cite{2023Kr03_PRC}.
\end{itemize}   
All these positive results are consistent with a production rate of $\approx 6 \times 10^{-6}$ times the corresponding nuclear reaction: $p\, + ^{Z}$X$ \,\to\, ^{Z+1}$Y$ \: + \: \gamma$.

However, these results have not been confirmed by an independent experiment. The CERN NA64 experiment, employing high energy bremsstrahlung $e^{-} Z \to e^{-} X$ reactions, found no evidence of anomalous $(e^{+}e^{-})$ production at 17 MeV \cite{2020Banerjee_PRD}. The observed anomaly cannot be explained within the Standard Model without stretching parameters to unrealistic values \cite{2021Aleksejevs_arXiv} and numerous theoretical investigations have proposed new physics models to explain the anomaly  \cite{2022Viviani_PRC, 2020Feng_PRD, 2021Zhang_PLB, 2016Feng_PRL, Boehm:2002yz, Boehm:2003hm, Alves}. 
There is general agreement among experimental and theoretical communities on the urgent need for a new independent experiment. We aim to meet this challenge by building a state-of-the-art detector with world-first capabilities and performing the experiments outlined below to test these anomalies and significantly improve on the current experimental sensitivity .

\section{Low Energy Nuclear Reactions as a Probe of New Physics}

\begin{figure}
 
   \vspace*{-14pt} \hspace*{0pt}
   \centerline{ 
    \resizebox{0.9\textwidth}{!}{\includegraphics{{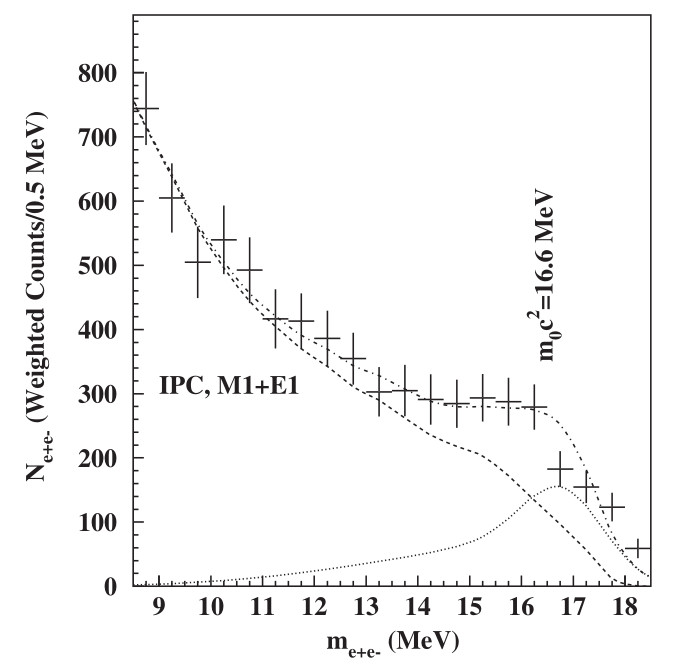}} \hspace*{36pt} \resizebox{0.9\textwidth}{!}{\includegraphics{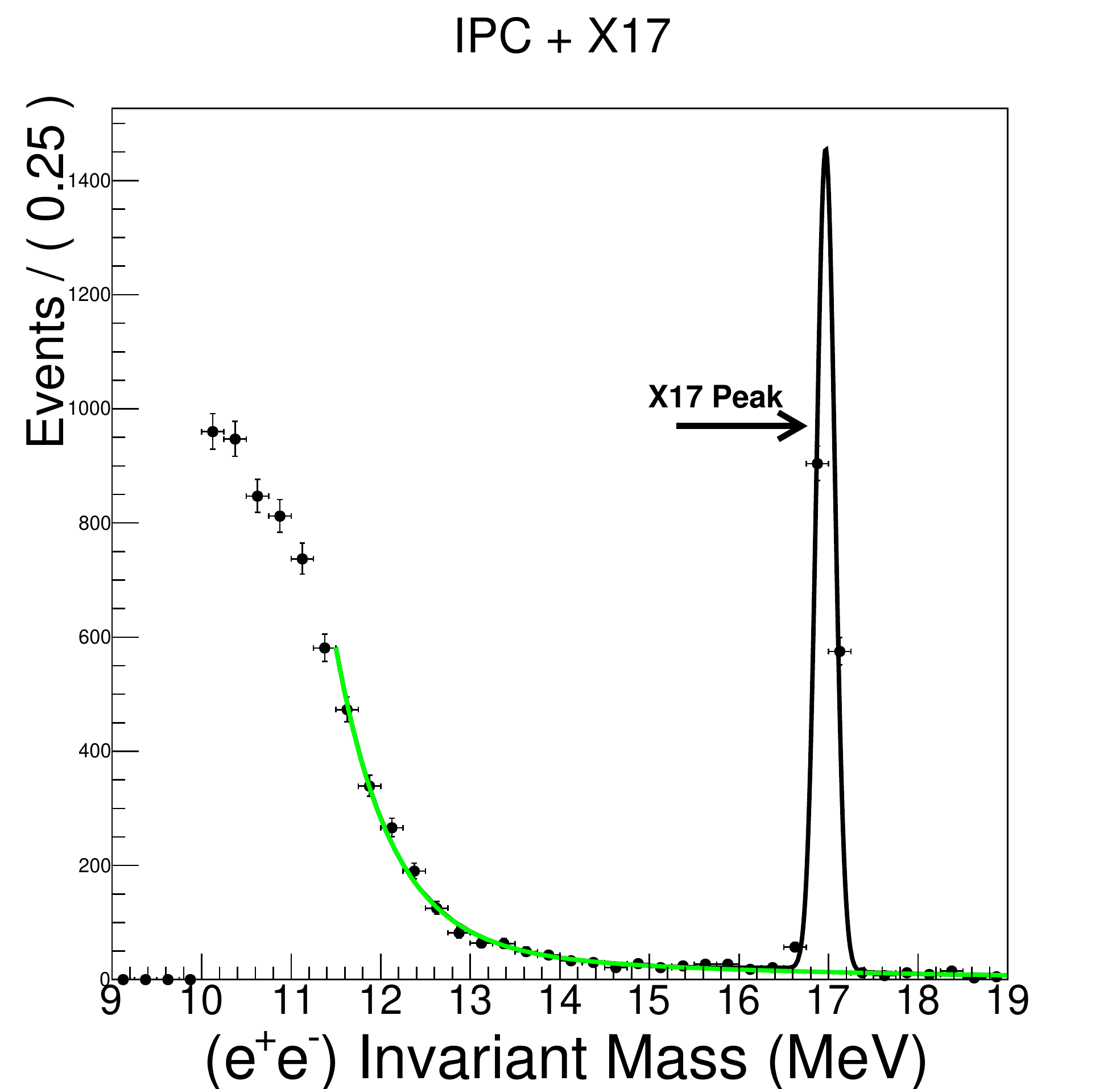}}}}

  \caption{Left is figure 5 from \cite{2016Kr_PRL}  which shows the invariant mass distribution of the $e^{+}e^{-}$ measured by the ATOMKI group. The dotted curve peaked at 16.6 MeV is the proposed contribution of the X17 convoluted with the experimental resolution. The anomaly differs from expectations by $\approx 7 \sigma$. Right are simulations of the $p + ^{7}$Li$ \,\to\, ^{8}$Be$\: +( e^{+}e^{-}) $ reaction for the TPC after 4 days of running on the Pelletron assuming X17 production at the rate found by the ATOMKI group together with the expected IPC background. The signal due to the X17 is the peak at 17 MeV. The green line is a fit to an ansatz which models the IPC contribution.}
  \label{fig:TPC_prod}
\end{figure}

There are now numerous projects to search for the X17 using high-energy particle physics experiments. However, no searches which employ the same nuclear reaction which observed the anomaly have yet been made, other than by the original experimenters. We intended to employ the University of Melbourne 5 MV Pelletron accelerator to initiate the  $p + ^{7}$Li$ \,\to\, ^{8}$Be$\: +( e^{+}e^{-}) $ reaction and to build a low mass, high precision Time Projection Chamber (TPC) to provide a far more sensitive search for the X17 and to search for any other anomalous $(e^{+}e^{-})$ yield. If the X17 exists as an independent fundamental particle, it will be evident as a peak at the corresponding invariant mass of the $e^{+}e^{-}$ final state. Furthermore, the TPC will be employed to explore a larger mass region (5 - 21 MeV) and to search for new physics with 2 orders of magnitude greater precision than has been achieved to date. The irreducible background is Internal Pair Conversion (IPC), where a virtual photon facilitates a nuclear decay and subsequently converts to an $(e^{+}e^{-})$ pair. This produces a broad background peaked at low invariant mass and decreases exponentially to the kinematic limit. Our instrument is therefore designed to have the best resolution possible in the invariant mass of the $(e^{+}e^{-})$ pair. \Cref{fig:TPC_prod} shows the expected performance of our TPC relative to the original X17 anomaly and the ease with which we will observe the X17 if produced at the rate found by the ATOMKI group. Note that the TPC has over an order magnitude better invariant mass resolution.

 Previous experiments using nuclear transitions that show evidence for the X17 boson
 were all carried out by the ATOMKI group \cite{2016Kr_PRL,2021Kr02_PRC,2023Kr03_PRC} who employed the $p + ^{7}$Li$ \,\to\, ^{8}$Be$\: +( e^{+}e^{-}) $, $p + ^{3}$H$ \,\to\, ^{4}$He$\: +( e^{+}e^{-}) $ and $p + ^{11}$B$ \,\to\, ^{12}$C$\: + (e^{+}e^{-})$ reactions.  In the case of $p + ^{7}$Li$ \,\to\, ^{8}$Be$\: +( e^{+}e^{-}) $, there are two spin-parity $1^{+}$ states at 17.64 and 18.15 MeV excitation energy in $^{8}$Be, which can be selectively populated using resonances at 0.441 MeV and 1.100 MeV in the proton beam energy. ATOMKI found an $\approx 7\sigma$ excess of events at high separation angles from the $E_{p} = 1.1$ MeV run but not at $E_{p} = 0.441$ MeV~\cite{2016Kr_PRL}. Subsequent measurements of the  $p + ^{3}$H$ \,\to\, ^{4}$He$\:+ (e^{+}e^{-}) $ reaction at incident proton energies of $E_{P} =0.510, 0.610,$ and $0.900$ MeV ($E_x$= 20.21, 20.29 and 20.49 MeV) found an $\approx 7 \sigma$ excess at all three energies\cite{2021Kr02_PRC}.
 Their setup, based on scintillator and position-sensitive detector arrays around the target, 
 has high efficiency but has no provision to select $e^{-} \, e^{+}$ pair events over other types of radiation.
 As a result, the yield of the X17 decays was only a tiny fraction, far less than 1 in a million,
 of the total yield. Furthermore, the invariant mass resolution of their setup fundamentally limits the precision with which they can search for other signals of new physics in these nuclear reactions.
 
 We propose to construct a TPC which provides magnetic selection and accurate particle tracking to overcome the limitations of the previous experiments. The TPC has large acceptance, excellent background rejection, and vastly improved invariant mass resolution which enables significantly more sensitive searches. In addition, the TPC allows us to make accurate measurements of the angular distributions of the $e^{+}e^{-}$ particles, enabling us to determine the spin and parity of the final state and hence that of any hypothetical boson.
 
 Our initial plan is to investigate the $p + ^{7}$Li$ \,\to\, ^{8}$Be$\: +( e^{+}e^{-}) $ reaction with 2 orders of magnitude higher sensitivity than ATOMKI, running at the same energy. Following this we will run at higher energies to search for evidence of new physics which could potentially be unveiled by the unprecedented sensitivity of our detector.
 
\begin{figure}

   \centering
    \resizebox{0.9\textwidth}{!}{\includegraphics{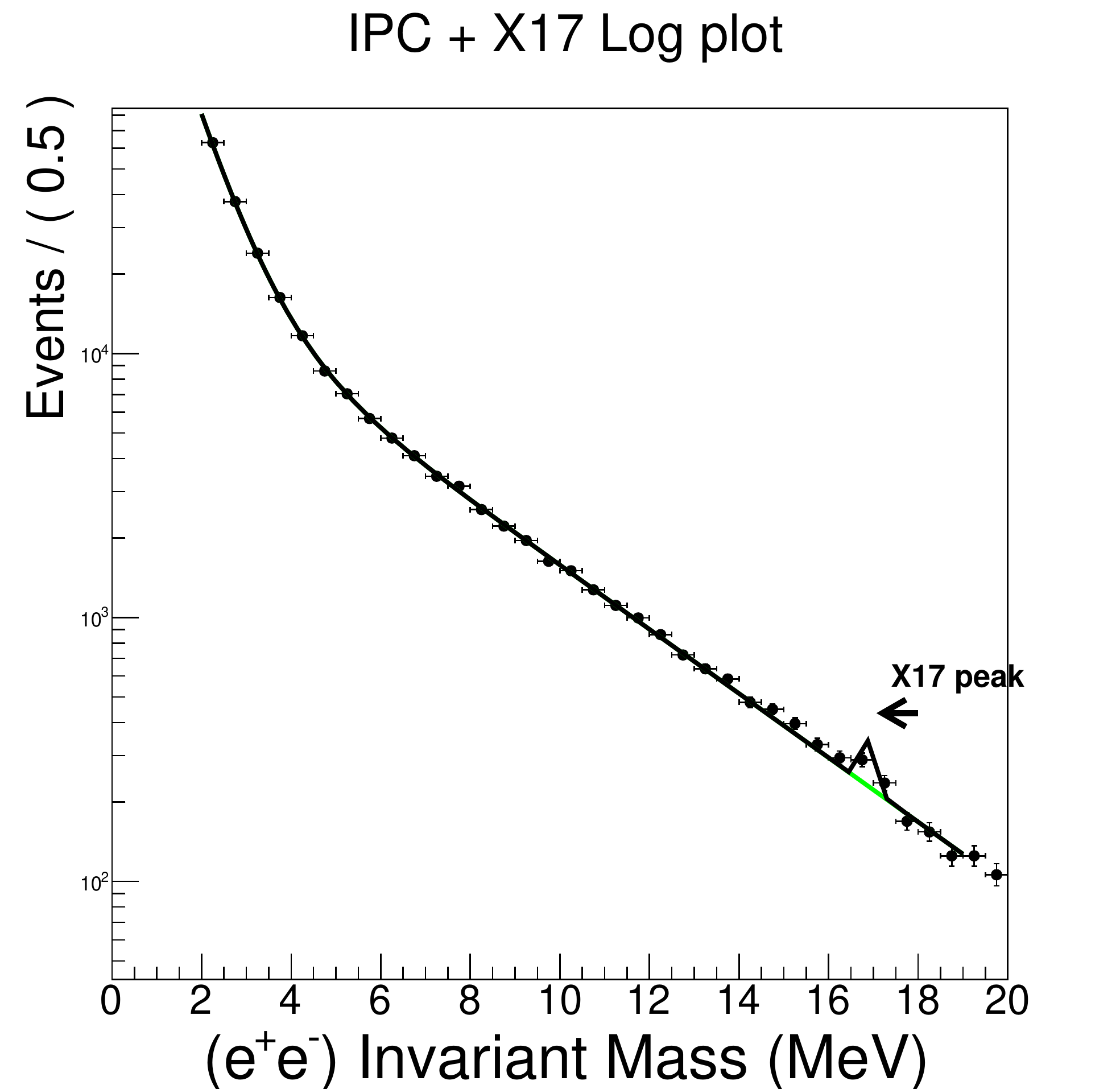}}
  \caption{Simulations of signal and background for the $p + ^{7}$Li$ \,\to\, ^{8}$Be$\: +( e^{+}e^{-}) $ reaction for the TPC after 30 days of running on the Pelletron, assuming an X17 produced at a rate 200 times smaller than  the ATOMKI observation \cite{2016Kr_PRL}. 
  }
  \label{fig:TPC_reach}

\end{figure}

Assuming a target thickness of the $2\times10^{19}$ atoms/cm$^2$ and a proton beam current of 2 microamp, the results of a 4-day run on the Pelletron are shown in \cref{fig:TPC_prod} for an X17 production rate of $6\times 10^{-6}$ that of $p+^{7}L \,\to\, ^{8}B + \gamma$, as found by Krasznahorkay et al.~\cite{2016Kr_PRL}. If it exists at the rate observed by the ATOMKI group, we will observe it with {\bf greater than 100 $\sigma$ significance} over a 30-day run (see \cref{fig:TPC_fits}). Otherwise the 30-day run on the Pelletron will place upper limits of the production of new physics signals in these reactions over {\bf two orders of magnitude smaller}, allowing us to place significantly stronger bounds on new physics models as shown in \cref{fig:TPC_reach}.

\section{Proposed Time Projection Chamber}

\begin{figure}
   \centerline{ 
    \resizebox{0.95\textwidth}{!}{\includegraphics{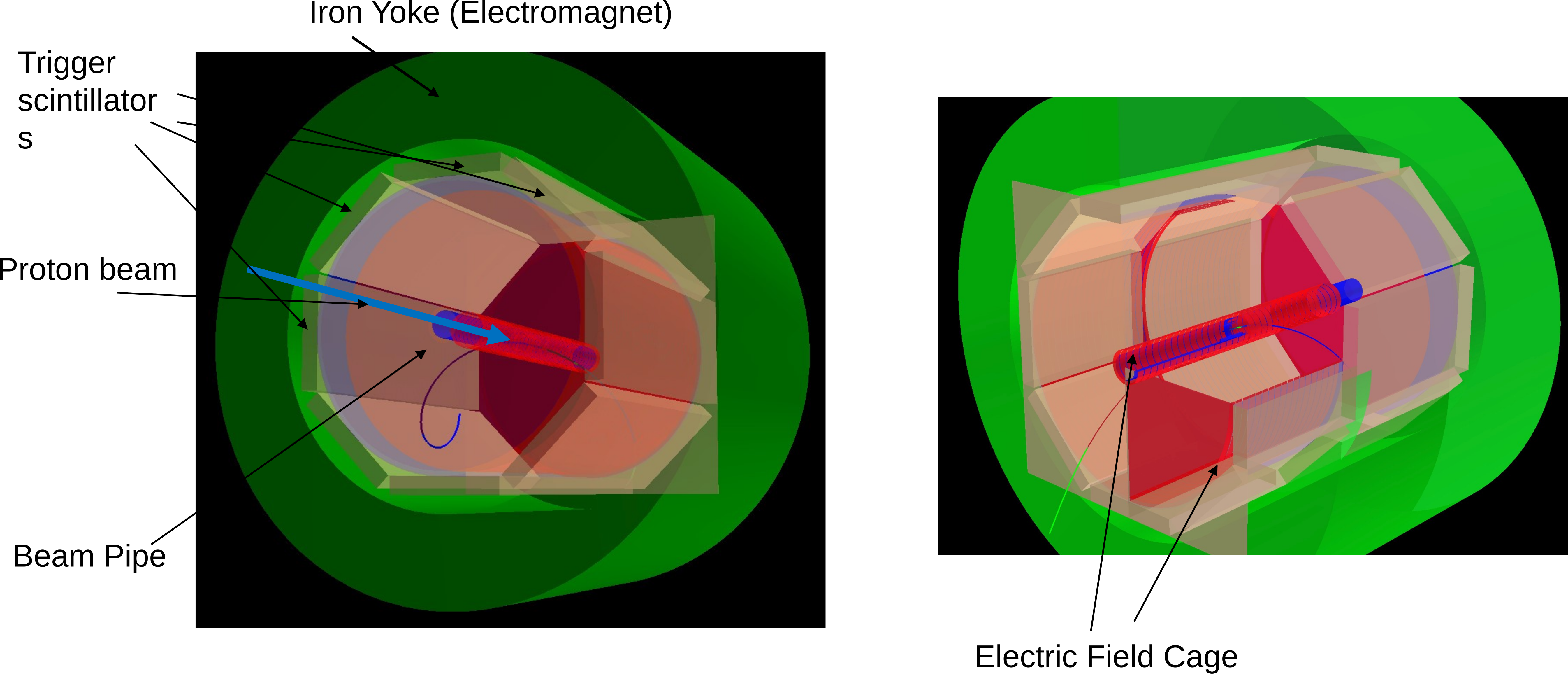}}}
  \caption{Overall design of the TPC. The internal diameter of the magnet yoke is 40 cm. The TPC is instrumented to a radius of 17 cm. The proton beam is directed down a vacuum tube of 1.0 cm radius of a target located in the center of the detector. Simulated $(e^{+}e^{-})$ pairs originate from nuclear reactions on the target. The right figure is a cut-away view showing the field cage of the TPC.}
  \label{fig:TPC_overall}
\end{figure}

The TPC provides 3-dimensional tracking of charged particles 
by reconstructing the ionization path of their passage through a gaseous medium. As charged 
particles (in this case the $e^{+} \, e^{-}$ pairs) traverse the medium, they ionize the gas and liberate electrons. 
The medium is placed inside parallel electric and magnetic fields. The magnetic field is employed to determine the momentum of the charged tracks.
The conceptual design of the TPC was made using the Event Visualization Environment (EVE) event-display package \cite{EVE} within the HEP-Physics root \cite{root} framework. The overall arrangement of the detector is shown in \cref{fig:TPC_overall}.

The TPC is a low-mass device made with low-$Z$ materials. 
 The outer shell will be constructed from a 1 mm thick Aluminium sheet.  Inside the outer shell, we place an electric field cage to provide a uniform electric drift field to guide liberated electrons to the anode plane. The inner wall of the TPC consists of an Aluminum plated Kapton sheet held at ground potential located at a radius of 1.5 cm from the proton beam. The volume inside the electric field cage is isolated from the outer volume of the TPC. This outer volume is filled with pure CO$_2$ gas. The inner volume contains the active gas mixture of the device (90:10 He to CO$_{2}$ by volume). This arrangement minimizes contamination of the sensitive gas region from oxygen, water vapor, and other impurities found in the air. The CO$_2$ is also an excellent electrical insulator that prevents internal sparking. Holding the entire outer shell at ground potential significantly improves electrical safety and serves to minimize stray electromagnetic interference. Just before the anode readout plane, we place a Micromegas stage \cite{micromegas}.

The proposed TPC-based detector system consists of a magnetic solenoid with a 40 cm internal 
diameter, which provides magnetic fields of up to 0.4 Tesla. 
Inside is an arrangement of 8 scintillators which detect the $e^{+} \, e^{-}$ pairs 
from the X17 decay X$17\to e^{+}e^{-}$. 
The TPC is placed within the scintillator array and covers radii from $r=1.5$ cm to 17 cm. 
Proton beams are transported through a 1 cm radius diameter beam pipe and impinge on 
the target placed in a target chamber at the center of the TPC. The construction of the target chamber is optimized to minimize multiple scattering, which limits the invariant mass resolution of the $e^+ e^-$ pair.

\begin{figure}
   \centerline{ 
    \resizebox{0.9\textwidth}{!}{\includegraphics{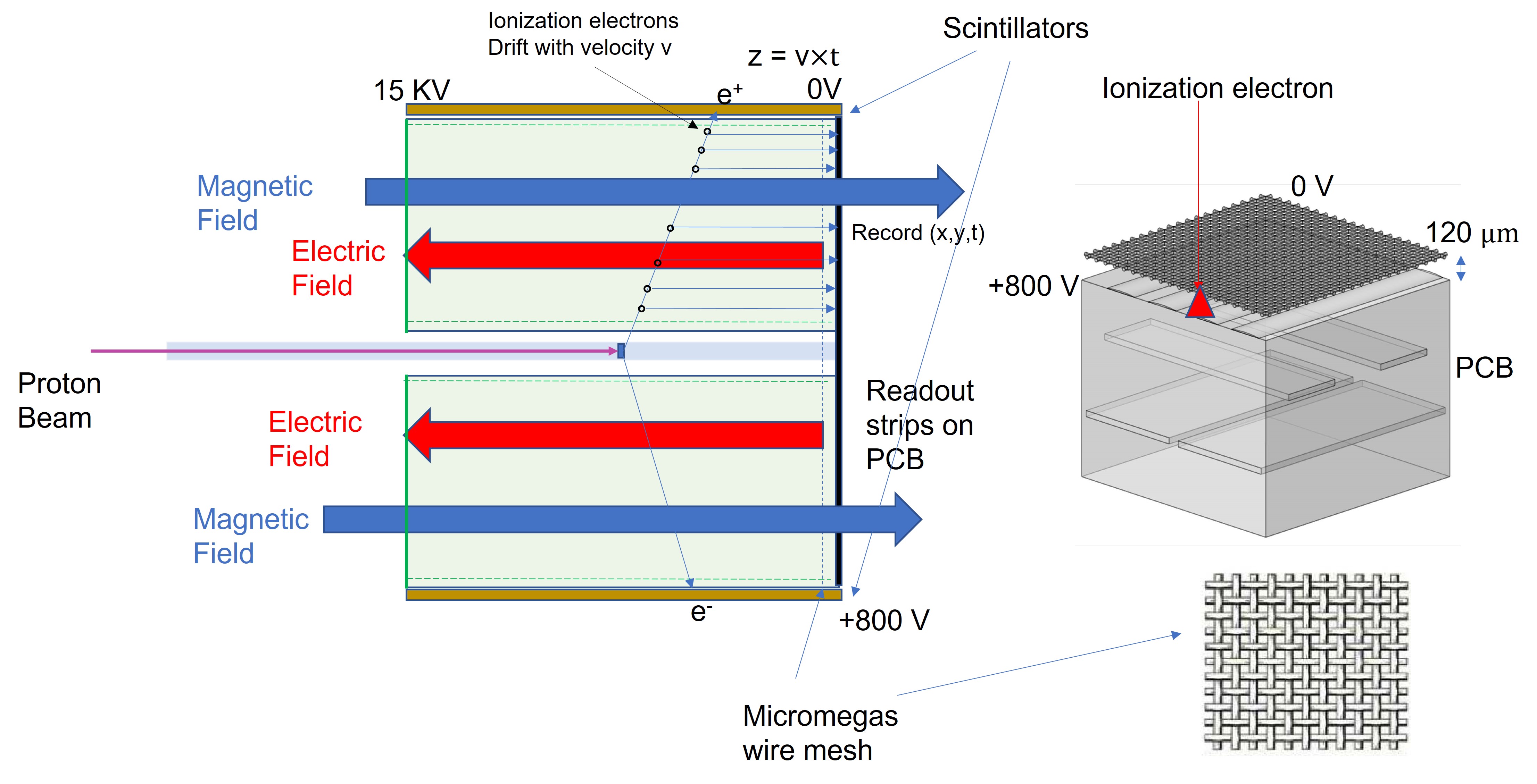}}}
  \caption{A side-on view of the TPC concept. The details are explained in the text. }
  \label{fig:TPC_concept}
\end{figure}

\begin{figure}
  \centerline{ \resizebox{0.9\textwidth}{!}{\includegraphics{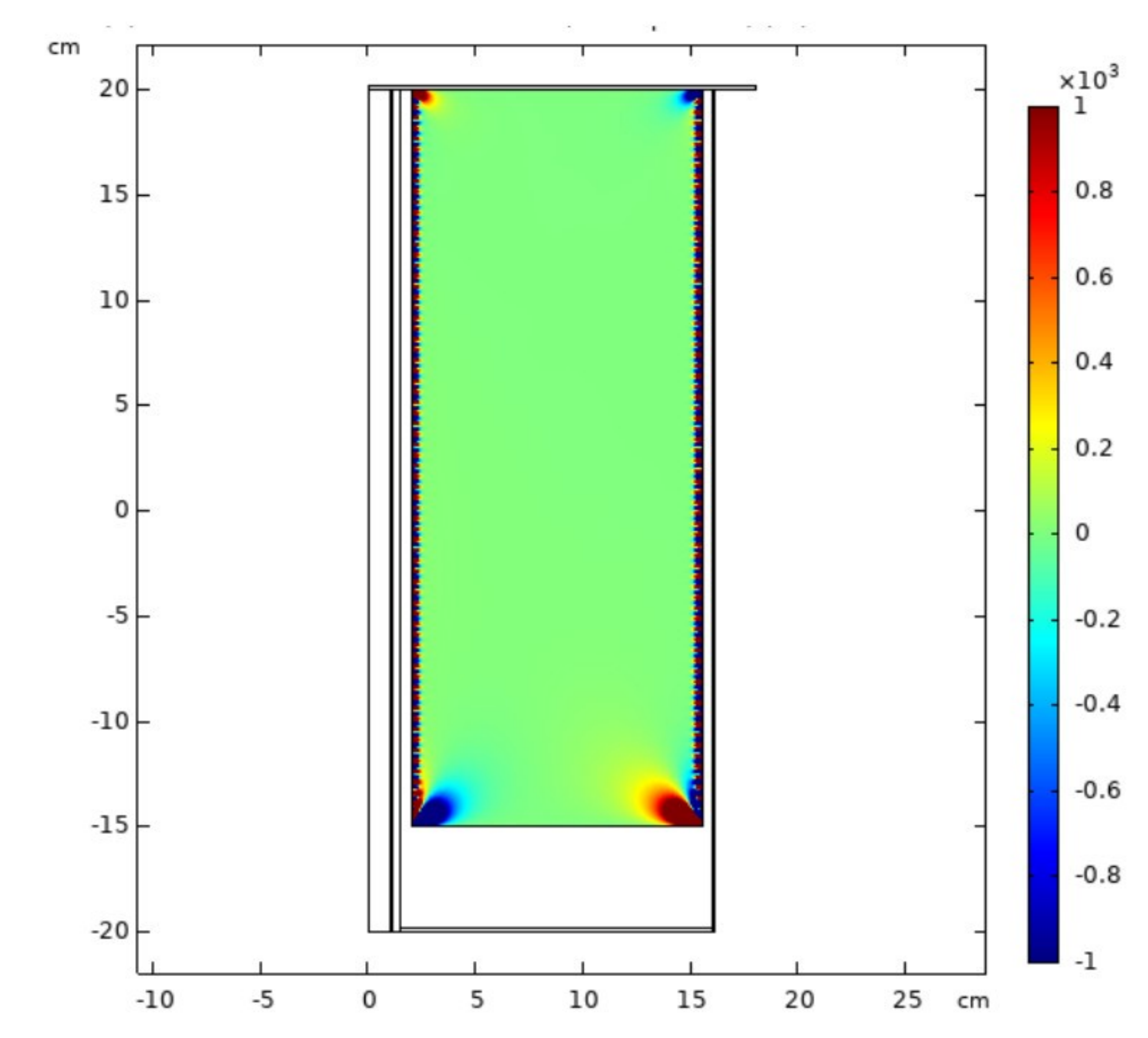}}}
\caption{COMSOL simulation of the radial electric field in the drift region of the TPC.}
\label{fig:TPC_Z_field}
\end{figure}

The conceptual operation of the TPC is shown in \cref{fig:TPC_concept}. A uniform electric field provided by high voltage stepped down from -15 KV to 0 V over the 35 cm length of the active region. This was achieved via an electric field cage of aluminized rings on the inner and outer walls of the TPC. A uniform magnetic field is provided by an electromagnetic solenoid. The high energy $e^{+}e^{-}$ pair from nuclear reactions have a radius curvature inversely proportional to their momentum. The TPC is filled with a He/CO$_2$ gas mixture in a ratio of 90:10 at atmospheric pressure. As the $e^{+}e^{-}$ traverse the detector, they ionize the gas and liberate electrons. These drift along electric field lines at constant velocity until they reach the Micromegas gas amplification region. This amplifies the electron signal by a factor of $\approx 10^4$. These induce pulses on the X-Y readout strips located in a multilayered PCB. A coincidence between two of the scintillators is used to trigger the readout of the TPC and also to provide the time-zero to determine the drift time of the charged particle tracks through the device. Thus the time of arrival, as well as the $(x,y)$ location of the pulse, is recorded. This drift time and known electron drift velocity give the $z$-coordinate of the electron. The $(x,y,z)$ locations of the liberated electrons provide space-point measurements which are collected together to provide the 3-dimensional tracks of the $e^+e^-$ pair.

The electric fields within the TPC were modeled with the commercial multi-physics package COMSOL\cite{comsol} and an optimal solution was determined after a number of configurations were trialed. The upstream cathode is a 1mm thick PCB disk set to a potential of -15 KV. The downstream anode is set to zero volts on the Micromegas mesh positioned 125 microns above a multilayer PCB. The TPC drift field is defined as a series of concentric rings each 4.5 mm wide with a 0.5 mm gap between them. The potential stepped down from -15 V to 0 V via 1 M$\Omega$ resistors placed between each ring. The first ring is placed 1 mm downstream of the cathode. The field cage thus defined has a typical uniformity of 1 part in $10^3$. The COMSOL simulation is shown in \cref{fig:TPC_Z_field}.

\begin{figure}

   \centerline{ 
    \resizebox{0.9\textwidth}{!}{\includegraphics{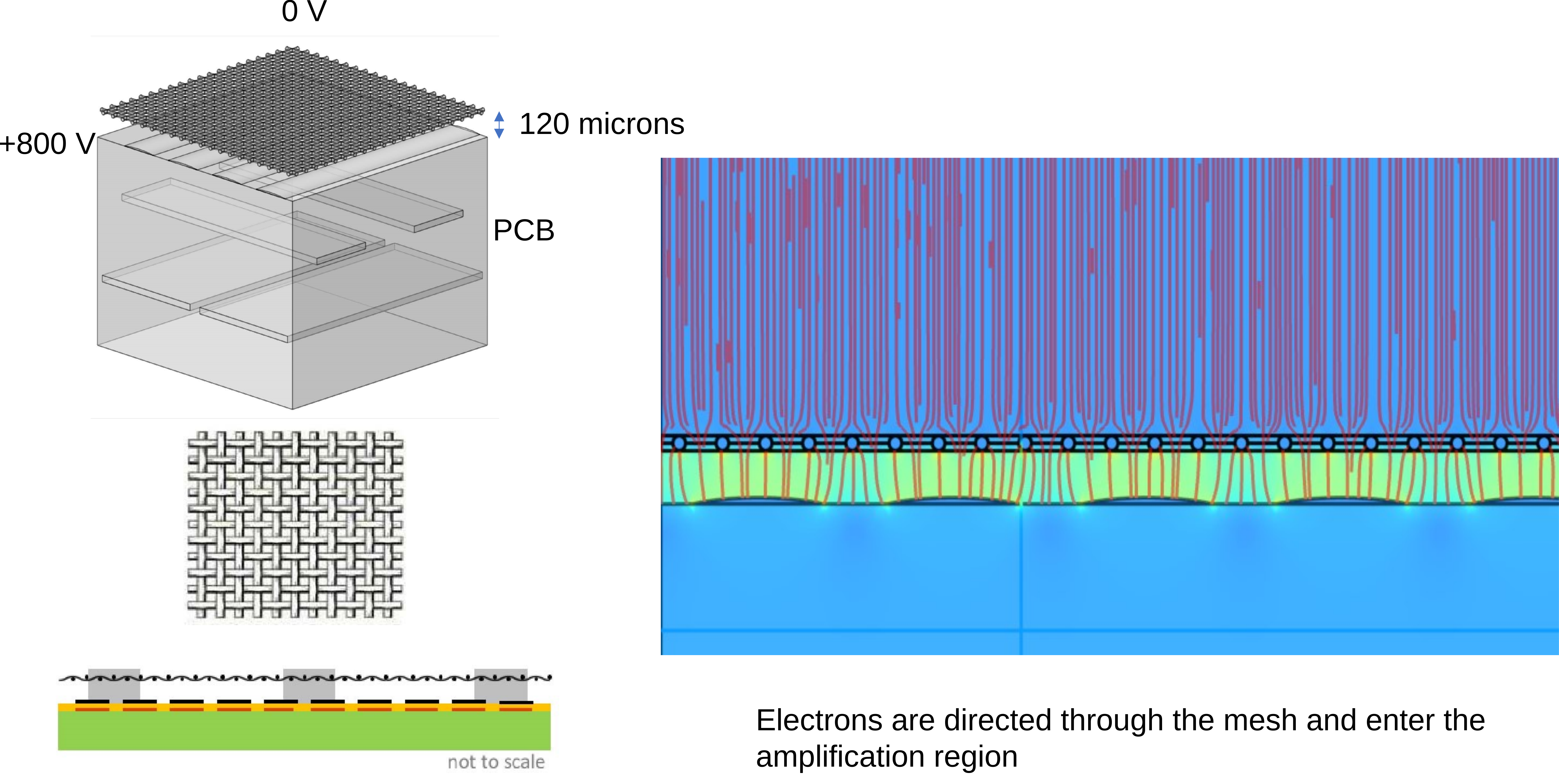}}}

  \caption{Left shows the concept of the Micromegas electron amplification stage consisting of a 15-micron pitch stainless steel wire mesh held 120 microns above 300-micron wide graphite strips on top of a multilayer PCB. The strips are held at 800 V. Subsequent layers hold X and Y strips of width 0.5 and 0.9 mm located 0.5 and 1 mm beneath the surface of the PCB. The mesh is held 125 microns above the surface with electrically insulating spacers. The image on the right shows the electron drift lines in the Micromegas electric field. This was also simulated by COMSOL. The electrons are directed around the mesh and into the gas amplification region. Further details are explained in the text. }
  \label{fig:TPC_micromegas}

\end{figure}

\begin{figure}

   \centerline{ 
    \resizebox{0.8\textwidth}{!}{\includegraphics{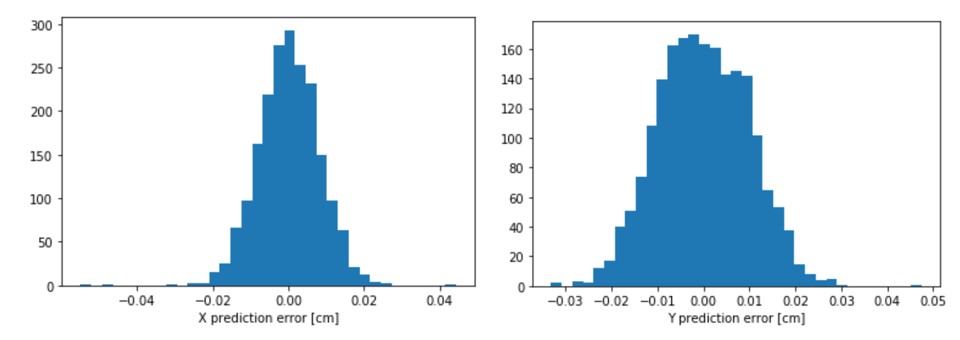}}}

  \caption{The X-Y resolution of the liberated electrons after projective readout onto the X-Y strips. The system provides $\approx 100\,\mu$m resolution}
  \label{fig:TPC_XY}

\end{figure}

\Cref{fig:TPC_micromegas} shows the layout of the micromegas amplification and readout region. We employ GARFIELD~\cite{garfield} to simulate the probability of electrons released liberated from the primary $e^{+}e^{-}$ via ionization of the He/CO$_2$ gas, the drift of electrons through the electric field, the Micromegas gas amplification, and the charge induced on the X-Y strips embedded in the PCB. The calculations predict electrons are liberated at a rate of $\approx$ 1/mm and drift with a velocity of 1 cm/$\mu$s with the -15 KV potential on the cathode. We employ the VMM ASIC~\cite{vmm} as the front-end chip to read out the strips. This provides 1 ns timing resolution as well as a 64 $\mu$s dynamic range to enable the full 35 $\mu$s readout time of the TPC. We employ the hit time on the strips to determine the drift time of the electrons as well enabling the association of the correct X-Y strips for projective readout of the Micromegas. The +800 V potential difference between the wire mesh and the graphite strips on the front of the PCB provides a gas gain of $\approx 10^4$ and results in induced pulses of $\approx$ 10 pC  spread across 2-3 strips in the X and Y planes. As shown in \cref{fig:TPC_XY}, taking the weighted mean of the distribution enables us to determine the location of the liberated electrons with a precision of $\approx 100\,\mu$m.

 \begin{figure}
   \centerline{ 
    \resizebox{0.48\textwidth}{!}{\includegraphics{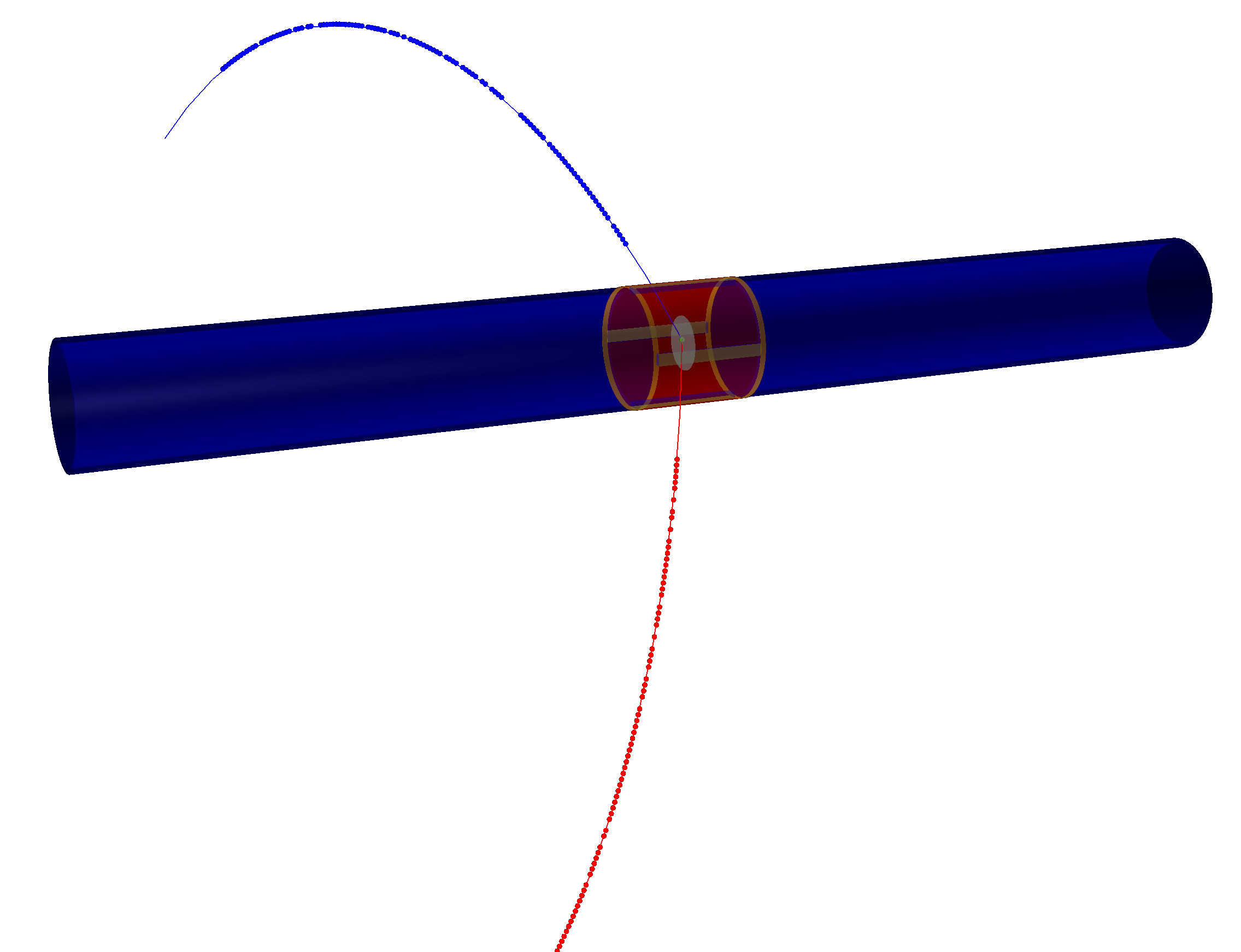}}
    \hspace*{6pt}
    \resizebox{0.48\textwidth}{!}{\includegraphics{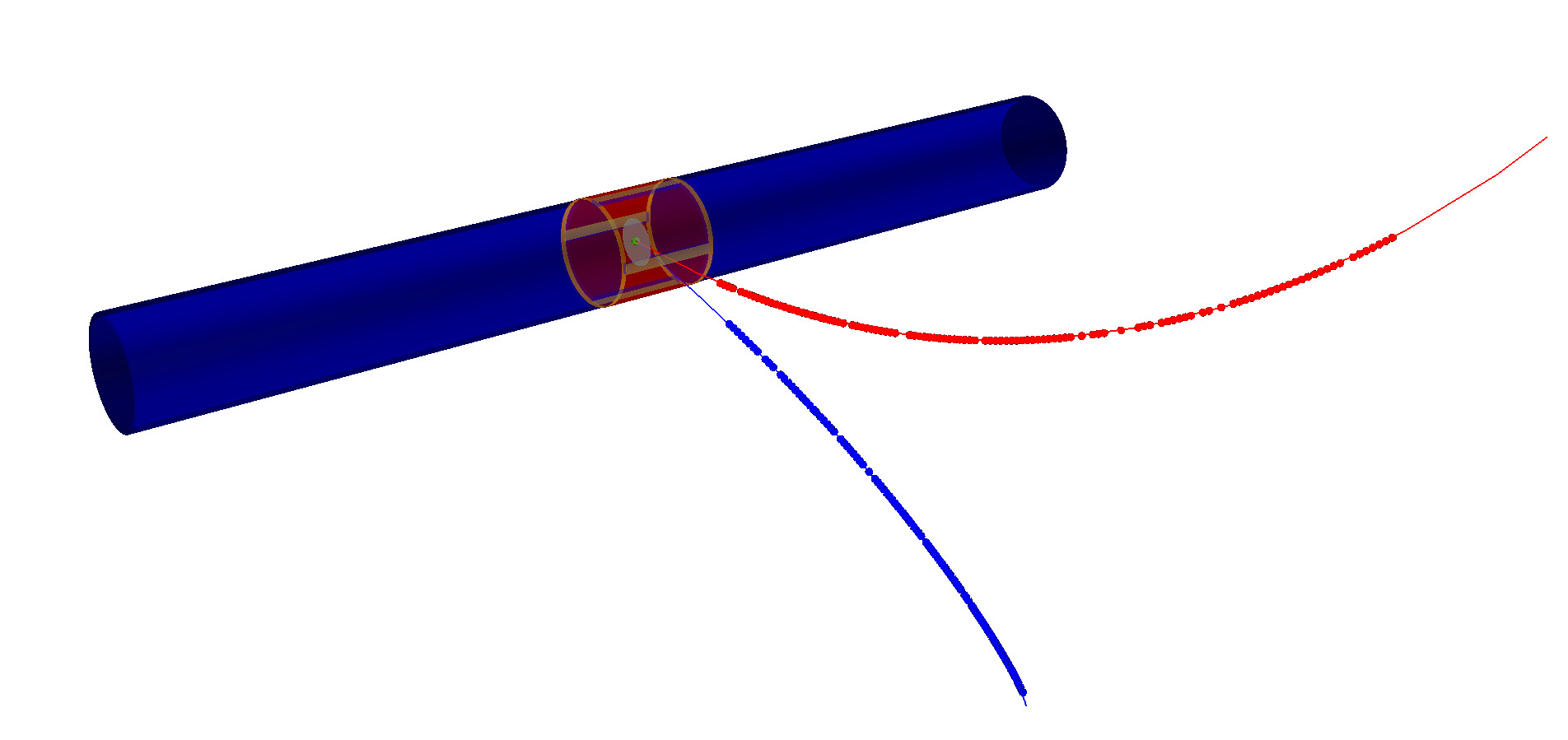}}}

      \caption{Event display of simulated typical X17 (left) and IPC (right) events. The reconstructed electron tracks are shown as blue lines and positrons are red lines. The dots show the location of the space points used to reconstruct the $e^{+}e^{-}$ trajectories. The tracks are projected onto the target and are required to form a common vertex constrained to originate on the target which is shown as a grey disk. In this event display, only the beam pipe and vacuum walls are shown. IPC events originate from off-shell photons and hence the invariant mass of such events is concentrated at low values. This results in a small opening angle between the $e^{+}e^{-}$ pair. In contrast, the X17 has an invariant mass near the kinematic limit leading to large opening angles for the $e^{+}e^{-}$ pairs.}
       \label{fig:TPC_events}

\end{figure}

\begin{figure}

   \centerline{ 
    \resizebox{0.475\textwidth}{!}{\includegraphics{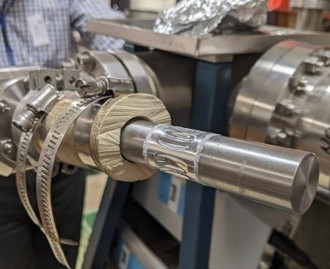}}
    \hspace*{6pt}
    \resizebox{0.48\textwidth}{!}{\includegraphics{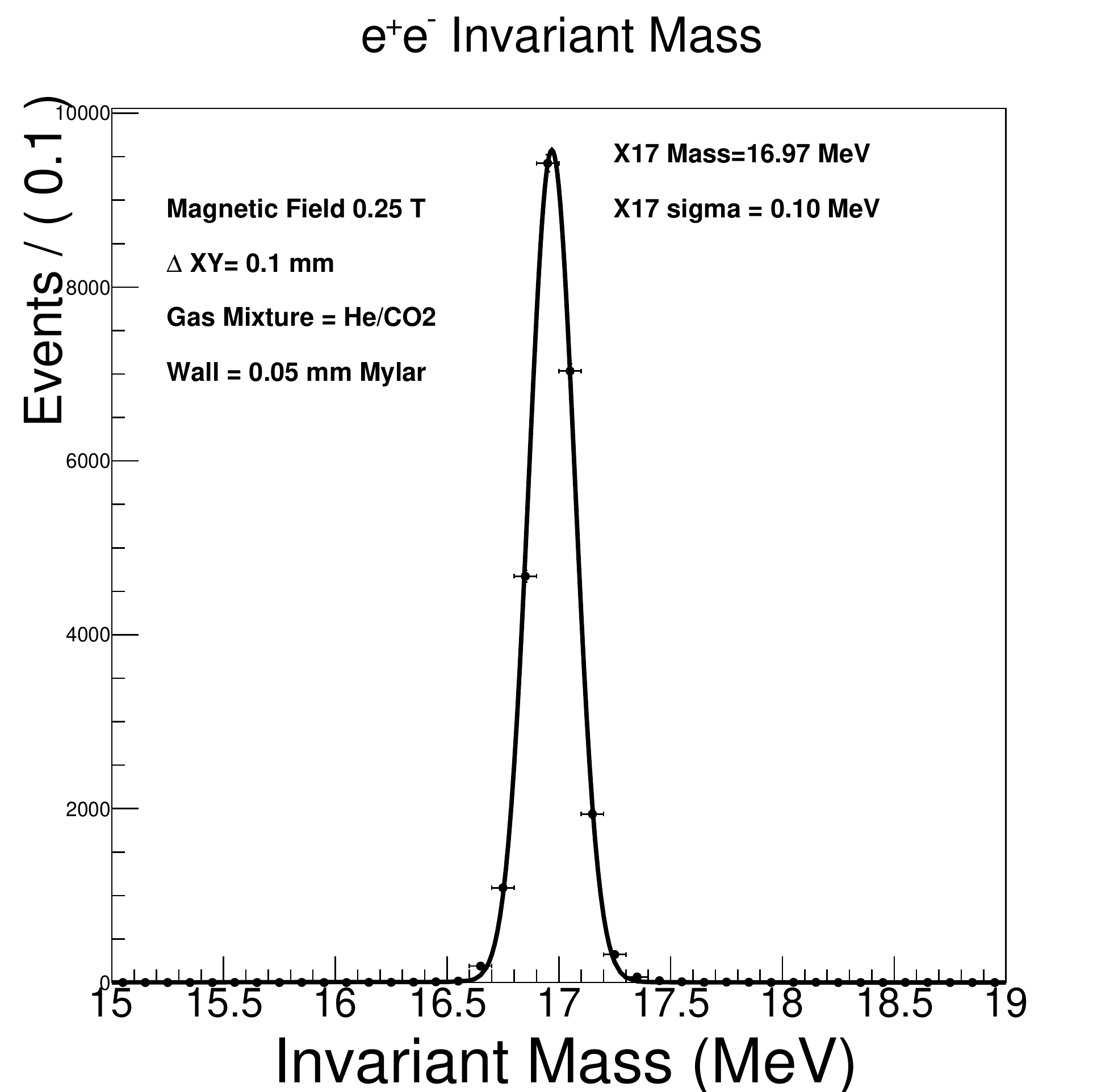}}}
      \caption{Left shows a photo of the prototype chamber under vacuum. Right shows the invariant mass resolution obtained from the full GEANT4 simulation using the thin mylar target wall, 90:10 He/CO$_2$ gas mixture, full reconstruction of the 130 hit collections per track, and vertex reconstruction.}
       \label{fig:chamber_res}

\end{figure}

We perform a full GEANT4 Monte-Carlo simulation of the TPC based on the Geometry Definition Markup Language (GDML) \cite{gdml} file generated from the EVE conceptual design. The model propagated the primary $e^{+}e^{-}$ pair through the TPC. It included energy loss and multiple scattering in the vacuum wall of the beam pipe, the inner walls of the TPC, and through the TPC gas-sensitive region itself. The electron ionization rate of 1 electron/mm in the He/CO$_2$ gas mixture results in $\approx$ 130 space points per track. These space points were reconstructed using the GENFIT2 \cite{GENFIT} Kalman-filter-based track-fitting package and the RAVE \cite{rave} vertex-finding package. \Cref{fig:TPC_events} shows event displays of simulated IPC and X17 processes. The invariant mass resolution of the TPC is limited by multiple scatting in the TPC gas and the vacuum chamber walls.  The reconstructed invariant mass was optimized by using a 90:10 He/CO$_2$ gas mixture and by employing a very thin vacuum wall consisting of 50 micron thick aluminized mylar. The sheer strength of Mylar is 15 kg/mm$^2$ which enables us to employ it as a very thin vacuum wall in a suitably designed target chamber. We made a prototype target chamber by milling four 2 cm long holes in an 11 mm radius cylinder of 2 mm thickness. This left four 2 mm wide posts to support the structure. Our calculations showed that this should provide a factor 20 safety margin against breaking under atmospheric pressure. The 50 micron thick mylar foil was glued over the vacuum pipe and posts using Torr Seal glue. We attached a turbo pump to the prototype and pumped it down. We found it to be mechanically stable and able to reach high vacuums. By placing cuts on the quality of the vertex fit we are able to select events where both tracks only pass through the mylar foil. Overall we find our overall invariant mass resolution for the X17 has a standard deviation of 0.1 MeV, which provides excellent discrimination against the smoothly varying IPC background. \Cref{fig:chamber_res} shows the prototype target chamber under vacuum and the expected X17 resolution with a magnetic field of 0.25 Tesla. After applying the reconstruction criteria, the overall acceptance of the TPC to the $p + ^{7}$Li$ \,\to\, ^{8}$Be$\: + X17 \to ( e^{+}e^{-})$ reaction is 17\% with a magnetic field strength of 0.25 Tesla.

Signals from the Micromegas are induced onto X-Y strips etched into a double-side circuit board. We will employ CERN-RD51 \cite{RD51} developed Scalable Readout Systems (SRS) components for our electronics. 
Pulses from the strips are fed into the Hybrid cards containing two 64-channel VMM  Application Specific Integrated
Circuit (ASIC) \cite{vmm} chips where they are amplified and digitized. Both pulse height and hit time are recorded. 
These data are read out via HDMI cables and sent through to the DVMM FPGA. This device enables sophisticated logic decisions while also providing a digital buffer 64 microseconds deep. This enables us to capture the time of arrival of the liberated electrons over the full drift time of the TPC. Data passing the DVMM trigger logic are transferred to the Front End Concentrator (FEC) which encodes the data into a 10 Gigabit ethernet stream for recording on a PC. This will be transferred to the University of Melbourne Spartan Research Computing platform for long-term storage and data analysis.

 The TPC enables a long-term program to search for new physics through nuclear reactions of the kind $p+ ^{Z}$X$ \,\to\, ^{Z+1}$Y$ + (e^{+}e^{-})$, as well as detailed studies of nuclear structure through $e^{+}e^{-}$ decays of excited states. The TPC will be constructed under contract by CERN in consultation with our team.

\section{Subdominant Backgrounds}

\begin{figure}

   \centerline{ 
    \resizebox{0.48\textwidth}{!}{\includegraphics{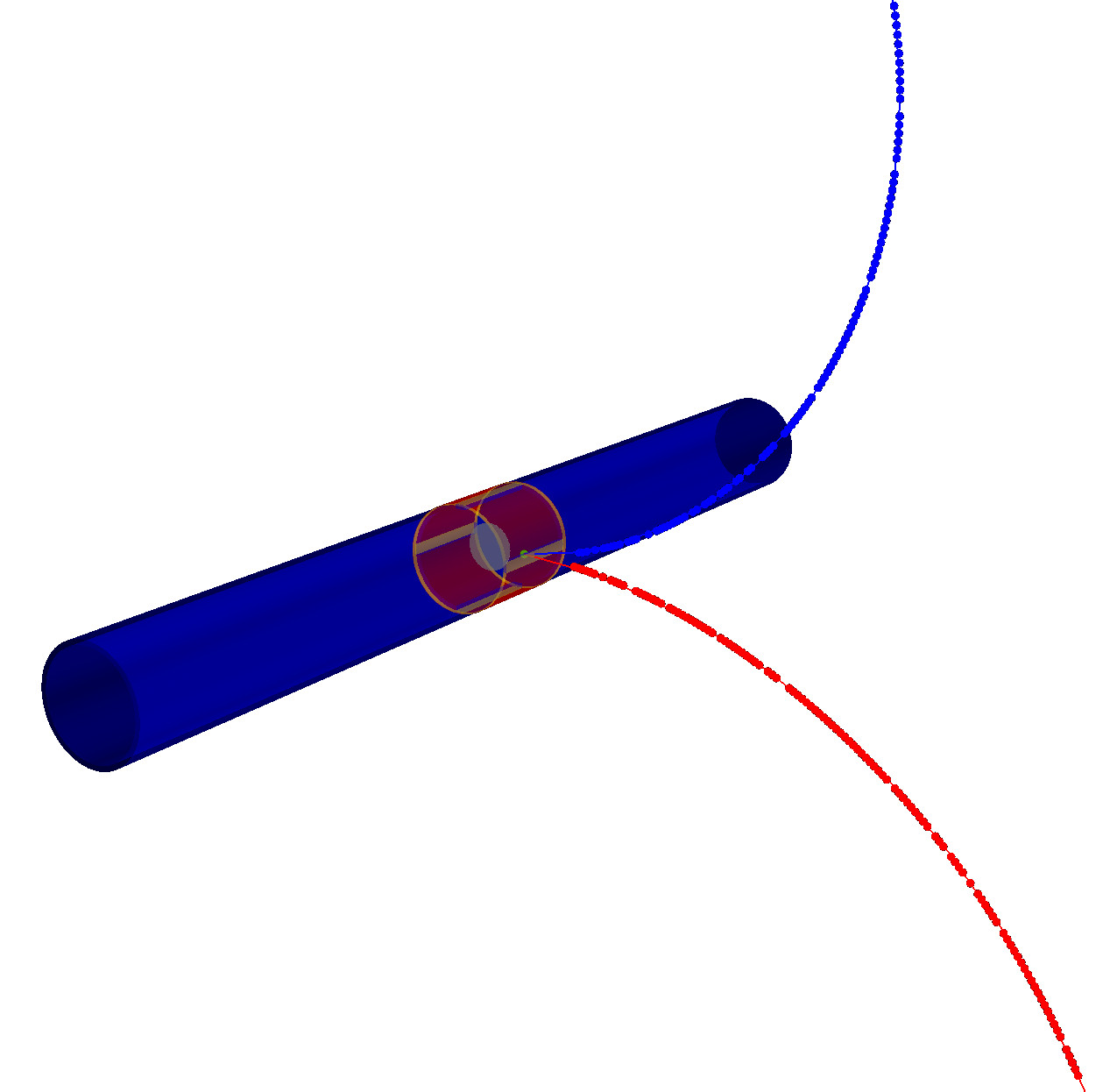}}
    \hspace*{12pt}
    \resizebox{0.48\textwidth}{!}{\includegraphics{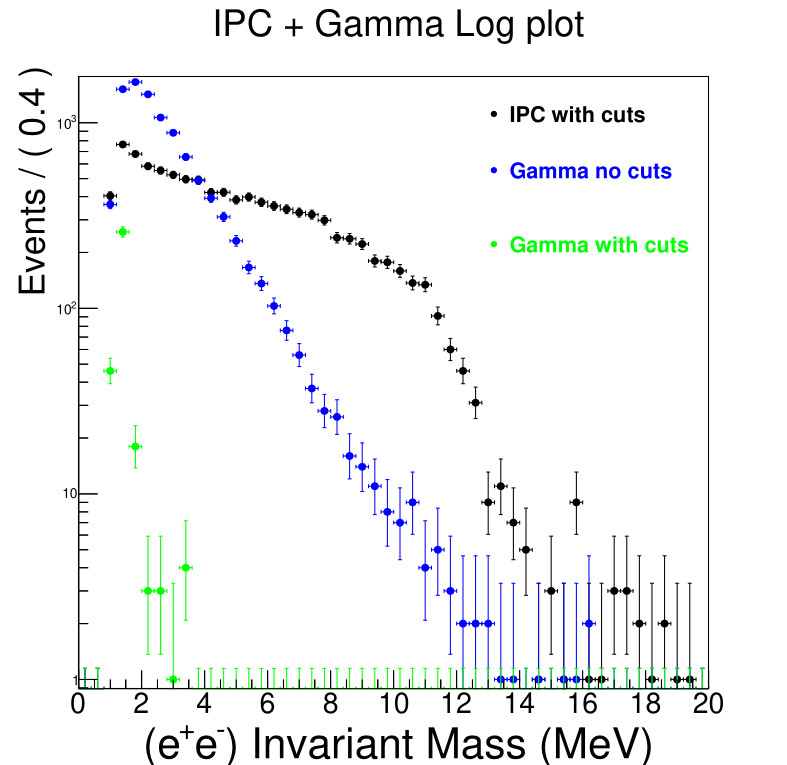}}}

  \caption{Left shows the simulation of typical  $p + ^{7}$Li$ \,\to\, ^{8}$Be$ + \gamma $ event detected by the TPC. The $(e^{+}e^{-})$ pair originates from a gamma conversion in the target wall. Such events are rejected by requiring the $e^{+}e^{-}$ to originate at a vertex on the target. The right plot shows the background from IPC and  $p + ^{7}$Li$ \,\to\, ^{8}$Be$ + \gamma $ events as a function the $e^{+}e^{-}$ invariant mass. The black points show the IPC background after applying cuts to ensure the $e^{+}e^{-}$ originates from a common vertex at the target. The blue points show simulations of the background from the  $p + ^{7}$Li$ \,\to\, ^{8}$Be$ + \gamma $ reaction without vertex constraints and the green points show the  $p + ^{7}$Li$ \,\to\, ^{8}$Be$ + \gamma $ background after applying the same cuts as the IPC data. The background from $p + ^{7}$Li$ \,\to\, ^{8}$Be$ + \gamma $ is less than 0.1\% that of IPC in the X17 mass region after applying vertex constraints. The distributions are those expected after a 10-hour run on the Pelletron.}
  \label{fig:gamma}

\end{figure}

As noted earlier, the irreducible background to the X17 and other new physics signals in proton-induced $e^{+}e^{-}$ reactions is the IPC process. The other large potential background is from the  $p + ^{7}$Li$ \,\to\, ^{8}$Be$\: + \gamma $ reaction, where the gamma subsequently undergoes external conversion via interaction with matter: $ \gamma \,\to\, e^{+} e^{-}$.  The $p + ^{7}$Li$ \,\to\, ^{8}$Be$ + \gamma$ reaction is $10^3$ times larger than IPC. To investigate this we simulated $10^{8}$ $p + ^{7}$Li$ \,\to\, ^{8}$Be$\: + \gamma $ reactions, approximately equivalent to 10 hours of running on the Pelletron. A typical event and the background expected from this process are shown in \Cref{fig:gamma}. After applying cuts to select $(e^{+}e^{-})$ with good vertices originating at the target, this background is reduced to less than 0.1\% that of IPC in the invariant mass region of the X17.

We considered cosmic rays and random beta decays as background. Primary cosmic ray muons in the $\approx$ 10 MeV/$c$ momentum region of X$17 \to e^{-}e^{-}$ have a kinetic energy of only 0.47 MeV so they do not have sufficient energy to initiate a trigger,  which requires 1 MeV energy loss in each scintillator. Random events require two beta decays each with $\sim$ 10 MeV kinetic energy with two reconstructed tracks within 5 mm on the target, with an invariant mass near the 17 MeV peak, and which fall within a ~10 ns time window. Consequently, both these backgrounds are expected to be vanishing small. Finally, although $\sim 9$ MeV $\alpha$ particles are copiously produced via the $p + ^{7}$Li$ \,\to\, \alpha + \alpha $ reaction,  they are all stopped in the inner detector walls and do not enter the sensitive region of the TPC.

\begin{figure}

   \centerline{ 
    \resizebox{0.48\textwidth}{!}{\includegraphics{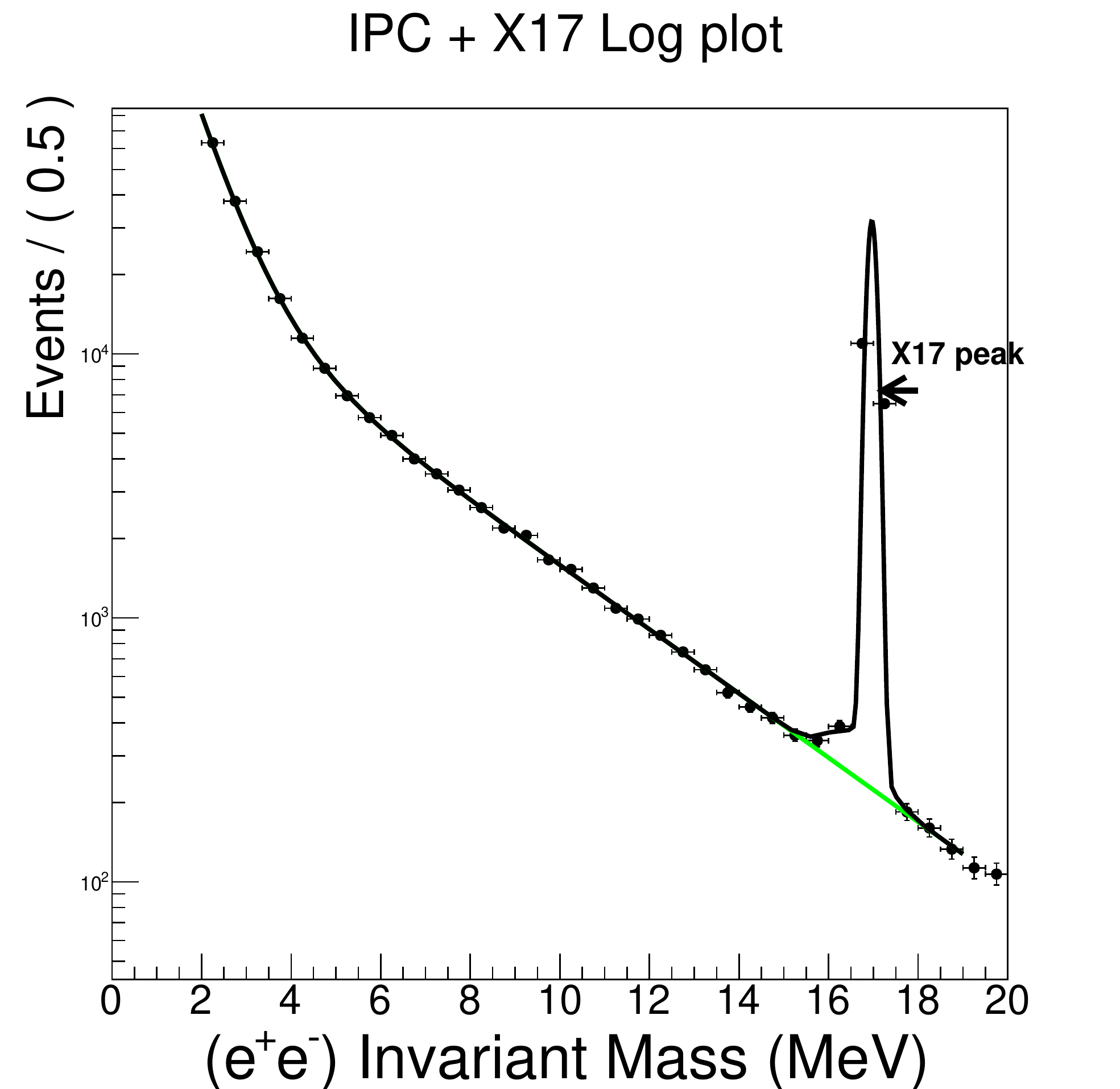}}
    \hspace*{16pt}
    \resizebox{0.48\textwidth}{!}{\includegraphics{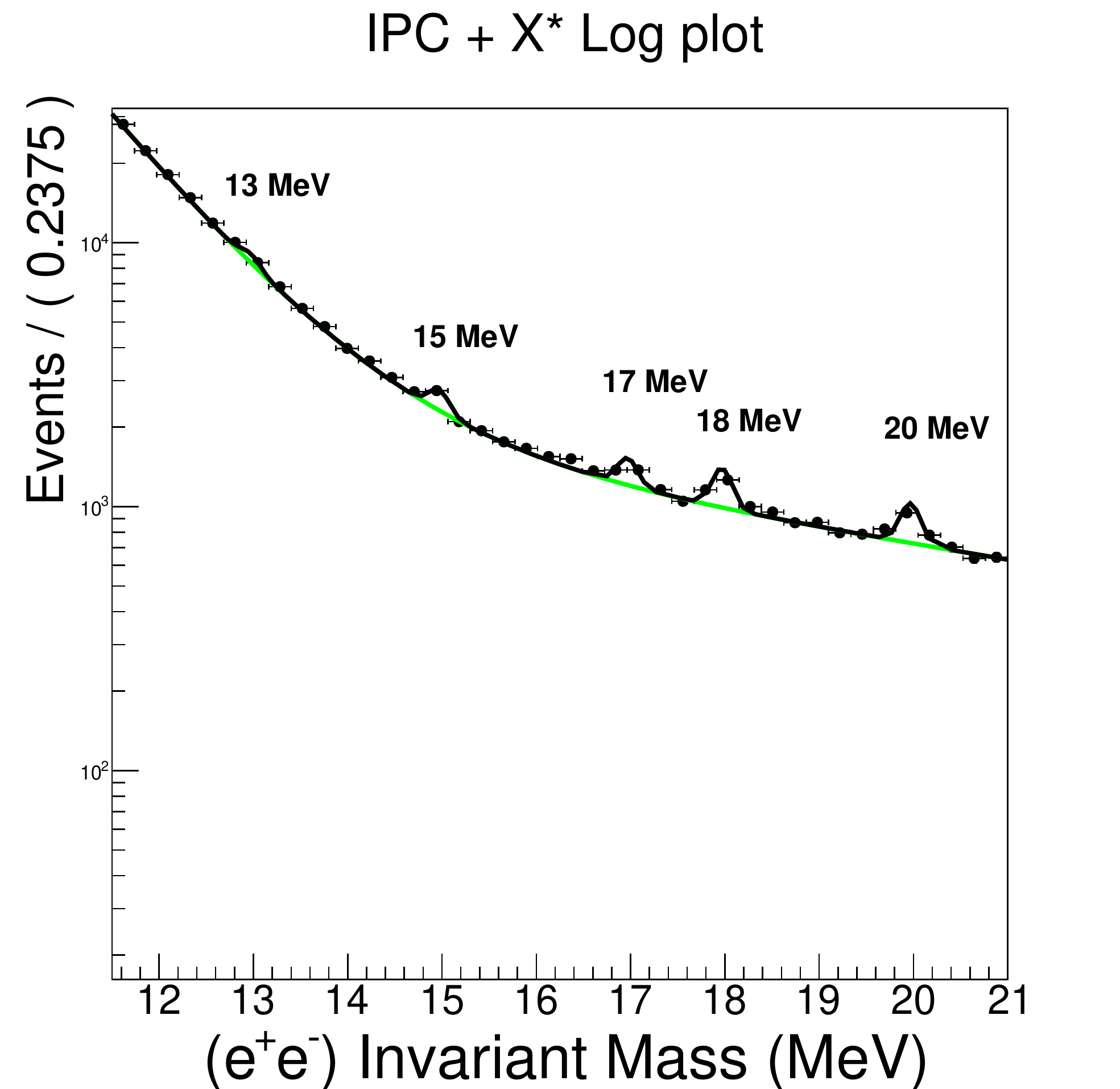}}}

  \caption{Left shows simulations of signal and background for the $p + ^{7}$Li$ \,\to\, ^{8}$Be$\: +( e^{+}e^{-}) $ reaction for the TPC after 30 days of running on the Pelletron assuming an X17 produced at a rate expected from the ATOMKI observation \cite{2016Kr_PRL}. Green is the double exponential ansatz used to model the background. Black is the sum of signal and background. We extract $17155 \pm 132$ events from the fit. Right shows simulations of a 30-day run at $E_p=4.5$ MeV, target thickness of $2\times10^{20}$ atoms/cm$^2$ and indicative results if feebly interacting bosons of 13, 15, 17, 18, and 20 MeV were produced at rates 200 times less than the X17.}
  \label{fig:TPC_fits}

\end{figure}

The background to X17 production is then overwhelmingly due to IPC events. \Cref{fig:TPC_fits} shows Roofit fits to simulations of the X17 signal and IPC background. The IPC  background is modeled with a two-component exponential ansatz while the X17 is modeled as Gaussian core plus a small tail from $(e^{+}e^{-})$ events that propagate through the edges of the thin mylar window.

\section{Experimental Research Programme}

We estimate the beam time required as follows. The total cross section for $p+ ^{7}$Li$ \,\to\, ^{8}$Be$ + \gamma$ at E$_{p} = 1.0$ MeV is $2.5\times 10^{-5} \: b$ \cite{Zahnow}. We assume that the $p+ ^{7}$Li$ \,\to\, ^{8}$Be$ + (e^{+} e^{-})$ IPC cross section is $3.6\times 10^{-3}$ times smaller. The Pelletron can comfortably supply beam currents of $ 2\mu A$ and we can make Li$_{2}$O targets of thickness $2\times 10^{19}$ Li atoms/cm$^2$. This corresponds to a proton beam energy loss of 0.040 MeV. The ATOMKI group found the branching ratio of X17 to the $p+ ^{7}$Li$ \,\to\, ^{8}$Be$ + \gamma$  reaction to be $6\times10^{-6}$ \cite{2016Kr_PRL}. Since the TPC has a 17\% acceptance for $p+ ^{7}$Li$ \,\to\, ^{8}$Be$ + $X$17 \to (e^{+} e^{-})$ we expect 17,000 X17 events in a 30-day run on the Pelletron if the ATOMKI group is correct. This situation is shown as the left plot in \cref{fig:TPC_fits}. Our simulations show that we would find $17155 \pm 132$ events, well over 100 $\sigma$ significance. We will also perform a `bump-hut' using the $E_{p} = 1.05$ MeV data to search for feebly coupled bosons. If none are found, we will set 90\% confidence upper limits as shown in Fig. \ref{fig:TPC_105_UL}. Depending on the results of the first runs we will either investigate the production of the X17 as a function of energy by running at E$_{p} = 1.5$, $2.0$ and $3.0$ MeV or make a high statistics run at E$_{p} = 4.5$ MeV to set the best upper limits possible. The cross section for $p+ ^{7}$Li$ \,\to\, ^{8}$Be$ + \gamma$ is similar ($3 \times 10^{-5} b$) at $E_{p}= 4.5$ \cite{fisher} and we will employ a target ten times thicker,  $2\times 10^{20}$ Li atoms/cm$^2$, corresponding to a proton beam energy loss of 0.11 MeV. This allows a greater reach in energy for feebly-interacting bosons and allows us to set strong limits on generic dark photons in the 8 -- 20 MeV mass range. We will then perform a bump hunt in the invariant mass distribution of the $e^{+}e^{-}$ pair. The plot on the right of \cref{fig:TPC_fits} shows simulations of the $E_{p}= 4.5$ MeV run. Here, we show indicative results of feebly interacting bosons of mass 13, 15, 17, 18, or 20 MeV decaying to $e^{+}e^{-}$ being produced with cross sections 200 times smaller than the ATOMKI X17.

The 90\% confidence levels were determined via toy Monte-Carlo experiments where we search for upward fluctuations of the background that mimic signal. For these we employ roofit to simulate 1000 experiments of pure background, generated randomly using the PDF fit to our GEANT4 simulations of the IPC process. We then search for signal as a function of mass, where the expected signal PDF is also determined by GEANT4 simulations. The 90\% upper limit is the value that is greater than 90\% of the simulations.
%

\section{Expected Sensitivity}
%

 \begin{figure}
  \centering
  \includegraphics[height=0.9\textwidth]{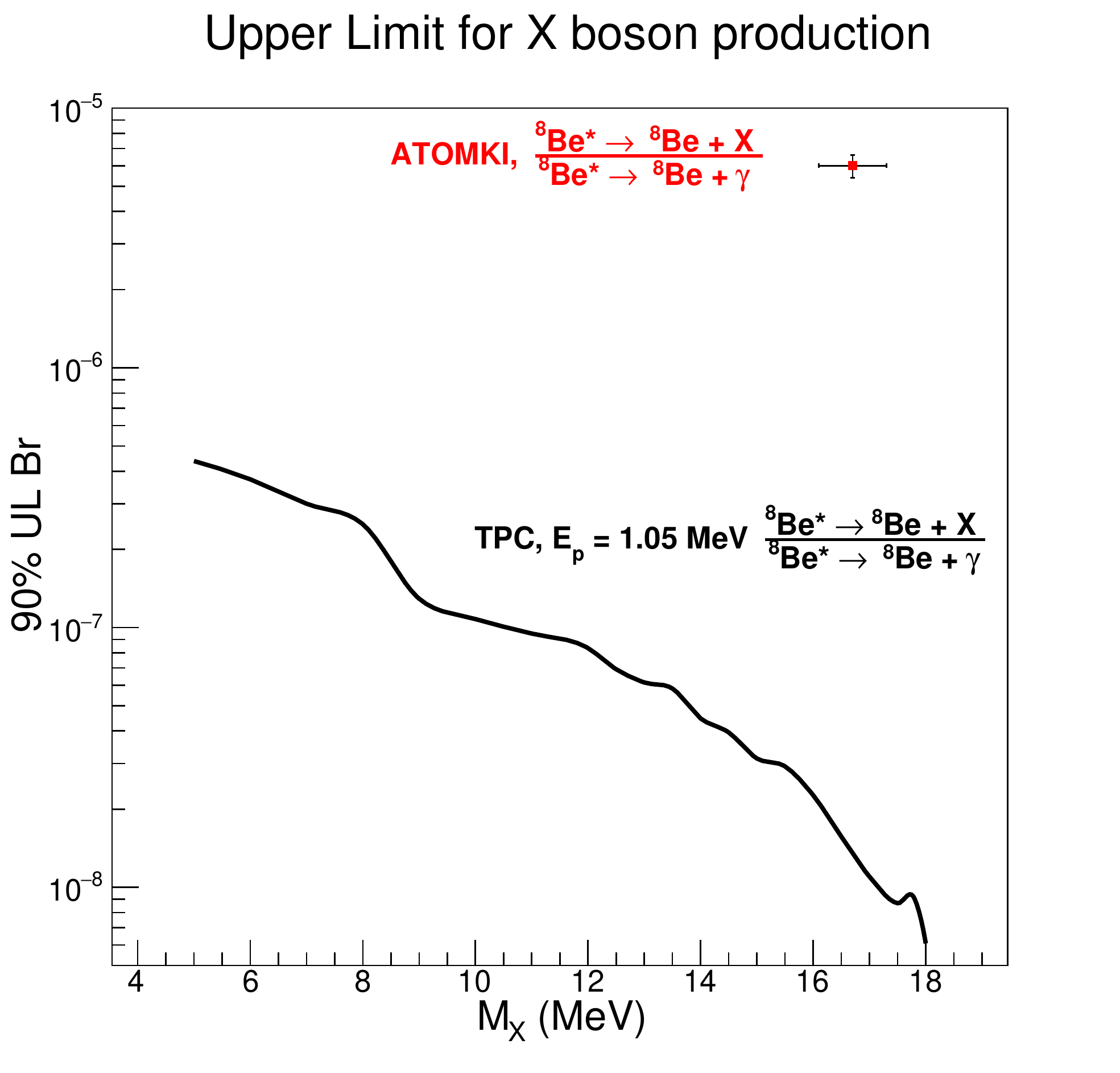}
  \caption{The black line shows the expected 90\% upper limits for a feebly interacting boson $X$ from the bump hunt at the TPC in the 30-day dataset for $p + ^{7}$Li$ \,\to\, ^{8}$Be$ +( e^{+}e^{-}) $ at $E_{p} = 1.05\,$MeV as a function of invariant mass of the boson. Also shown in red is the result for the ATOMKI experiment. The TPC will set an upper limit over two orders of magnitude smaller than ATOMKI at 17.0 MeV.  
  }
  \label{fig:TPC_105_UL}
\end{figure}
The proposed experimental research programme will conclusively test the $^8\text{Be}$ anomaly seen in the ATOMKI experiments and, in the absence of the discovery of a new particle, will provide world leading limits on a variety of important low energy nuclear reactions.  \Cref{fig:TPC_105_UL} shows the expected 90\% upper limit of the $^8\text{Be}$ search (solid black) in the mass range of $5 - 18 \,\text{MeV}$, along with the ATOMKI result at $E_p = 1.05$ MeV (red).  We see that the $^8\text{Be}$ run of the proposed TPC will probe the $^8\text{Be}^* \to {}^8\text{Be}(e^+e^-)$ branching ratio two orders of magnitude better than required to be sensitive to the ATOMKI $^8\text{Be}$ anomaly.

The origin of the ATOMKI Beryllium anomaly has been widely discussed in the literature. While \cite{2021Aleksejevs_arXiv} proposes a SM explanation, resonances beyond the SM have been explored in \cite{2022Viviani_PRC, 2020Feng_PRD, 2021Zhang_PLB, 2016Feng_PRL, Boehm:2002yz, Boehm:2003hm, Alves} (see also \cite{2017Feng_PRD} and references theirin).

\begin{figure}
  \centering
  \includegraphics[height=0.9\textwidth]{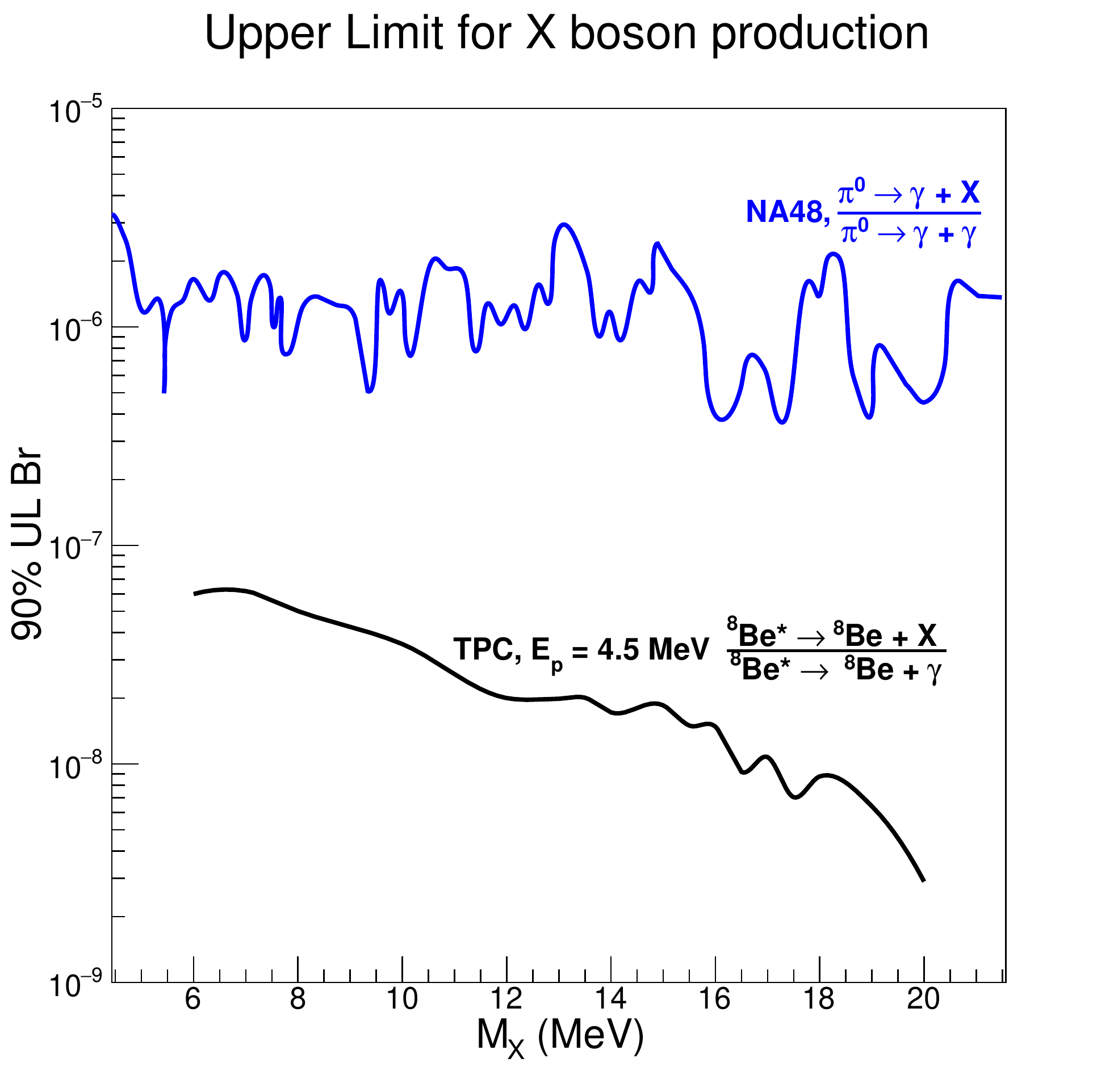}
  \caption{The black line shows the expected 90\% upper limits for a feebly interacting boson $X$ from the bump hunt at the TPC in the 30-day dataset for $p + ^{7}$Li$ \,\to\, ^{8}$Be$\: +( e^{+}e^{-}) $ at $E_{p} = 4.5\,$MeV. Also shown is the 90\% confidence upper limits from the NA48/2 experiment \cite{NA48} for feebly interacting boson $X$ from the $\pi^{0} \to \gamma + X(e^{+} e^{-})$ in the range $5 - 20\,$MeV. The TPC will set upper limits significantly below NA48/2 over the mass range 6 = 20 MeV. 
  }
  \label{fig:TPC_UL}
\end{figure}

Following an $E_p = 1.05\,$MeV run, a high statistics run at $E_p = 4.5\,$MeV will provide even stronger limits on the $^8\text{Be}^* \to {}^8\text{Be}(e^+e^-)$ branching ratio, particularly at lower masses, as shown in \cref{fig:TPC_UL}.

\subsection{Sensitivity to Dark Photons}

A dark photon is a hypothetical massive gauge boson \cite{Holdom:1985ag}, which is `dark' as it is related to a gauge symmetry in a hypothetical dark sector.  The dark photon can interact feebly with standard model particles due to kinetic mixing with the photon or through higher dimensional operators, and could act as a mediator between standard model particles and dark matter.\footnote{Higher dimensional operators (that is, operators with mass dimension greater than four) are non-renormalizable and require the introduction of additional fields.}  Assuming that the dark photon couples only via kinetic mixing, we can write the Lagrangian as~\cite{Fabbrichesi:2020wbt}
\begin{equation}  
\label{eq:L-dark-photon-1}
{\cal L} \supset 
- \frac{1}{4}\hat{F}_{\mu\nu}\hat{F}^{\mu\nu} 
- \frac{1}{4}\hat{F}^{'}_{\mu\nu}\hat{F}^{'\mu\nu} 
- \frac{\epsilon}{2}\hat{F}_{\mu\nu}\hat{F}^{'\mu\nu}
- \frac{1}{2}m^{2}_{X} \hat{A}^{'}_{\mu}\hat{A}^{'\nu} 
+ e\hat{A}_{\mu}\sum_{f}Q_{f}\bar{f}\gamma^{\mu}f \,,
\end{equation}
where $\hat{F}_{\mu\nu} = \partial_{\mu}\hat{A}_{\nu} - \partial_{\nu}\hat{A}_{\mu}$ and $\hat{F}^{'}_{\mu\nu}=\partial_{\mu}\hat{A}^{'}_{\nu}-\partial_{\nu}\hat{A}^{'}_{\mu}$ denote the field strength tensors of the photon and the dark photon before mixing, respectively, $\epsilon$ is the kinetic mixing parameter, $m_X$ is the mass parameter for the dark photon, $e$ is the unit of electric charge and $Q_f$ are the electric charges of the SM fermions $f$.  We can now perform the non-unitary transformation
\begin{equation}
\label{eq:dark-photon-transformation}
 \begin{pmatrix}
 \hat{A}_{\mu} \\
 \hat{A}^{'}_{\mu}
 \end{pmatrix}
 =
 \begin{pmatrix}
 1 & -\epsilon \\
 0 & 1
 \end{pmatrix}
 \begin{pmatrix}
 A_{\mu} \\
 A^{'}_{\mu}
 \end{pmatrix}\,,
\end{equation}
 to remove the kinetic mixing term. At order $\epsilon$ the Lagrangian, \cref{eq:L-dark-photon-1}, then becomes
\begin{equation}
\label{eq:L-dark-photon-2}
{\cal L} \supset 
- \frac{1}{4}F_{\mu\nu}F^{\mu\nu} 
- \frac{1}{4}F^{'}_{\mu\nu}F^{'\mu\nu} 
- \frac{1}{2}m^{2}_{A'}A^{'}_{\mu}A^{'\nu} 
+ e(\gamma_{\mu} - \epsilon A_{\mu})\sum_{f}Q_{f}\bar{f}\gamma^{\mu}f \,,
\end{equation}
where $F_{\mu\nu} = \partial_{\mu}A_{\nu} - \partial_{\nu}A_{\mu}$ and $F^{'}_{\mu\nu}=\partial_{\mu}A^{'}_{\nu}-\partial_{\nu}A^{'}_{\mu}$.

At the nucleon level the current $e\epsilon\sum_{f}Q_{f}\bar{f}\gamma^{\mu}f$ can be written as $e(\epsilon_{p}\bar{p}\gamma_{\mu}p + \epsilon_{n}\bar{n}\gamma_{\mu}n)$, where $p$ and $n$ denote the proton and neutron, respectively, $\epsilon_p = \epsilon(2Q_u + Q_d) = \epsilon$ and $\epsilon_n = \epsilon(Q_u + 2Q_d) = 0$. The $^8\text{Be}^* \to {}^8\text{Be}\,X$ branching ratio is~\cite{2016Feng_PRL}
\begin{equation}
\label{eq:dark-photon-br}
\frac{\text{BR}(^8\text{Be}^* \to {}^8\text{Be}\,X)}{\text{BR}(^8\text{Be}^* \to {}^8\text{Be}\,\gamma)} 
= (\epsilon_{p} + \epsilon_{n})^{2}\frac{|\vec{p}_{X}|^{3}}{|\vec{p}_{\gamma}|^{3}}
= \epsilon^{2}\frac{|\vec{p}_{X}|^{3}}{|\vec{p}_{\gamma}|^{3}}\,,
\end{equation}
where $\vec{p}_{X}$ and $\vec{p}_{\gamma}$ are the 3-momenta of the dark photon and the photon, respectively. 

 \begin{figure}
  \centering
  \includegraphics[height=0.65\textwidth]{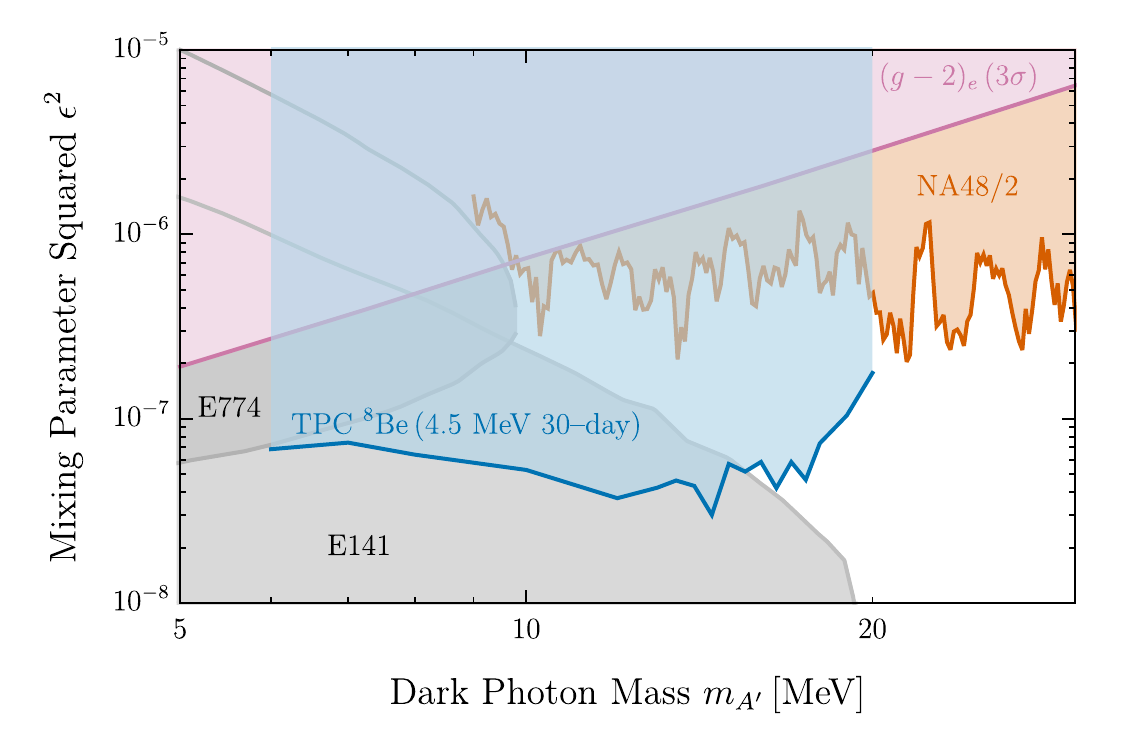}
  \caption{Constraints on the dark photon mixing parameter $\epsilon^{2}$ as a function of the dark photon mass from the TPC (solid purple line) together with constraints of the experiments NA48/2 \cite{NA48} and the beam dump experiments SLAC E141 and FNAL E774 \cite{E141}. Also shown are the limits derived from the measurement of the electron magnetic moment \cite{electron_g-2}. The result from the $E_{p} = 4.5$\,MeV run of the TPC provides the world's best exclusion for a promptly decaying dark photon in the $5-20$\,MeV range. 
  }
  \label{fig:TPC_eps}
\end{figure}

The upper limit shown in \cref{fig:TPC_UL} then provides limits on $\epsilon^{2}$, which is shown in \cref{fig:TPC_eps} together with the constraints of the experiments NA48/2~\cite{NA48}, the beam-dump experiments SLAC E141 and FNAL E774 \cite{E141} and the limits derived from the electron magnetic moment, $(g-2)_e$~\cite{electron_g-2}.
We see that the proposed TPC experiment will test an important unprobed region of parameter space in the mass range $6 - 20\,$MeV.

\subsection{Sensitivity to Axion-Like Particles}

A pseudoscalar explanation of the ATOMKI anomaly seen in the $^8\text{Be}(18.15)$ transition was first discussed in \cite{Ellwanger:2016wfe}. To investigate the TPC sensitivity to Axion-Like Particles (ALPs), we start with a Lagrangian describing the ALP coupling to nucleons \cite{Alves},
\begin{equation}
\label{eq:LALPnucleons}
{\cal L} = a \bar{N} i \gamma_5 (g_{aNN}^{(0)} + g_{aNN}^{(1)} \tau^3) N\,,
\end{equation}
where $N = (p, n)^T$ denotes the nucleon isospin doublet containing the proton and the neutron. The ALP coupling to the isosinglet and isotriplet currents is given by $g_{aNN}^{(0)}$ and $g_{aNN}^{(1)}$, respectively. 

The nuclear decay anomalies can be explained using ALPs if (i) the ALP mass $m_a$ in natural units is close to the invariant mass of the observed resonance $m_{ee}$ ($\approx 17$\,MeV) and (ii) the branching ratio for ALP decay to electron-positron pairs satisfies $\text{BR}(a\to e^+ e^-)\approx 1$, since the $a\rightarrow\gamma\gamma$ braching ratio is highly constrained in this mass region. Taking the Beryllium anomaly observed by ATOMKI as an example, the ratio of the ALP emission rate of $^8\text{Be}(18.15)$ to the corresponding photon emission rate is given by~\cite{Donnelly:1978ty, Barroso:1981bp, Alves}
\begin{align} \label{eq:BeWidth}
\frac{\Gamma(^8\text{Be}^*\to ^8\!\!\text{Be}+a )}{\Gamma(^8\text{Be}^*\to ^8\!\!\text{Be}+\gamma )}\approx\frac{1}{2\pi\alpha}
\left| \frac{g_{aNN}^{(0)}}{\mu^{0}-1/2}\right|^2\left(1-\frac{m_a^2}{\Delta E^2}\right)^{3/2}\,,
\end{align}
where $\alpha$ is the fine-structure constant, $\mu^0\approx 0.88$ is the isoscalar magnetic moment~\cite{Donnelly:1978ty} and $\Delta E=18.15$\,MeV is the energy splitting of the Beryllium transition. 

\begin{figure}
   \centering    \includegraphics[height=0.45\textwidth]{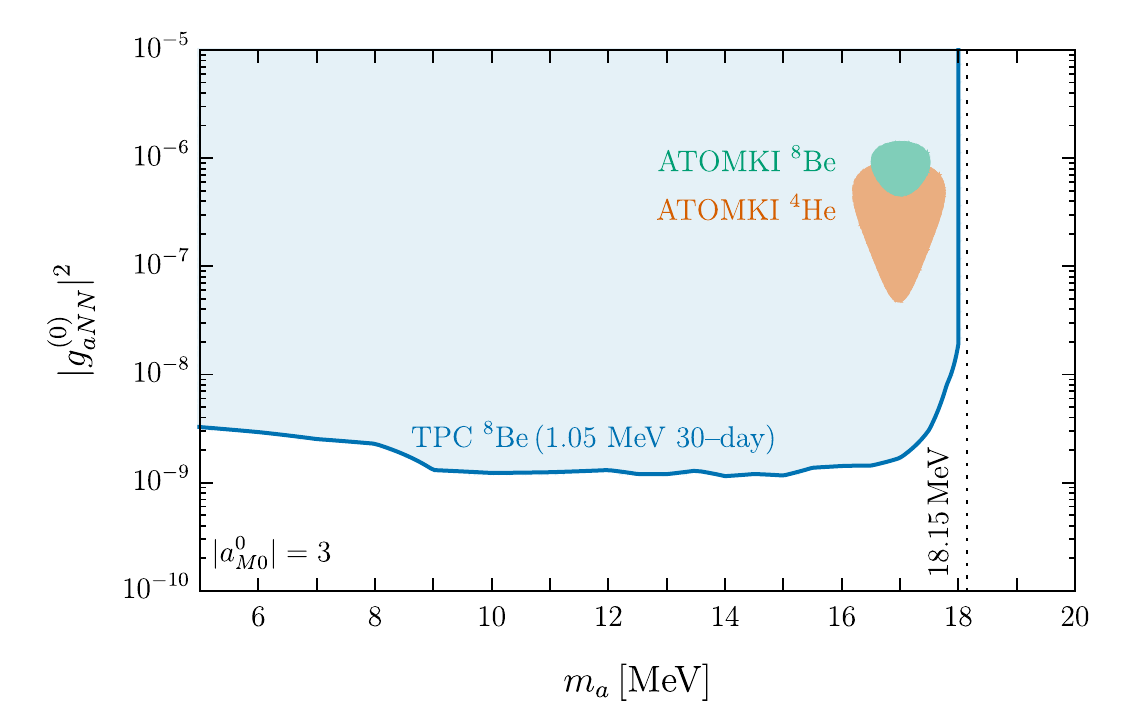}
  \caption{Reach of the proposed TPC experiment in the $m_a-|g_{aNN}^{(0)}|^2$ plane. The green (orange) regions are consistent with the ATOMKI observation in the $^8\text{Be}(18.15)$ ($^4\text{He}(21.01)$) transition, whilst the blue region would be excluded by the TPC if no resonance is observed.
 }
  \label{fig:ALPs-1}
\end{figure}

Assuming a null observation, the blue-shaded region in \cref{fig:ALPs-1} shows a projected exclusion limit from observations of the $^8\text{Be}(18.15)$ transition at the TPC in the $m_a-|g_{aNN}^{(0)}|^2$ plane, assuming the 90\% upper limit on the $^8\text{Be}$ branching ratio following the 30-day $E_p = 1.05\,\text{MeV}$ run as shown in \cref{fig:TPC_UL}. The orange region could explain the ATOMKI  $^8\text{Be}(18.15)$ measurement and the green region could explain the ATOMKI  $^8\text{He}(21.01)$ observation. The TPC experiment has the power to comfortably test these anomalies.

%
The ALP-nucleon couplings can be related to more fundamental ALP-SM couplings given at a high energy, or Ultra-Violet (UV), scale. These couplings are defined in terms of a general effective Lagrangian given here up to operators of mass dimension-5:
\begin{equation}\label{Leff}
\begin{aligned}
   {\cal L}_{\rm eff}^{D\le 5}
   &= \frac12 \left( \partial_\mu a\right)\!\left( \partial^\mu a\right) - \frac{m_{a,0}^2}{2}\,a^2
    + \frac{\partial^\mu a}{f}\,\sum_F\,\bar\psi_F\spac\bm{c}_F\,\gamma_\mu\spac\psi_F 
    + c_\phi\,\frac{\partial^\mu a}{f}\, 
    \big( \phi^\dagger i \hspace{-0.6mm}\overleftrightarrow{D}\hspace{-1mm}_\mu\spac\phi \big) \\
   &\quad\mbox{}+ c_{GG}\,\frac{\alpha_s}{4\pi}\,\frac{a}{f}\,G_{\mu\nu}^a\,\tilde G^{\mu\nu,a}
    + c_{WW}\spac\frac{\alpha_2}{4\pi}\,\frac{a}{f}\,W_{\mu\nu}^A\,\tilde W^{\mu\nu,A}
    + c_{BB}\,\frac{\alpha_1}{4\pi}\,\frac{a}{f}\,B_{\mu\nu}\,\tilde B^{\mu\nu} \,,
\end{aligned}
\end{equation}
where $f = -2 c_{GG} f_a$ is related to the ALP decay constant $f_a$ and defines the new physics scale $\Lambda = 4 \pi f$ which is the scale at which some new global symmetry is spontaneously broken, with the ALP $a$ emerging as the associated pseudo Nambu-Goldstone boson. $G_{\mu\nu}^a$, $W_{\mu\nu}^A$ and $B_{\mu\nu}$ are the field-strength tensors of $SU(3)_c$, $SU(2)_L$ and $U(1)_Y$, the dual field strength tensors are labelled with a tilde in the usual way, and $\alpha_s=g_s^2/(4\pi)$, $\alpha_2=g^2/(4\pi)$ and $\alpha_1=g^{\prime\,2}/(4\pi)$ denote the corresponding couplings. The sum in the first line extends over the chiral fermion multiplets $F$ of the SM and the quantities $\bm{c}_F$ are $3\times 3$ hermitian matrices in generation space. The field $\phi$ represents the Higgs doublet.

\begin{figure}
   \centering
    \includegraphics[height=0.45\textwidth]{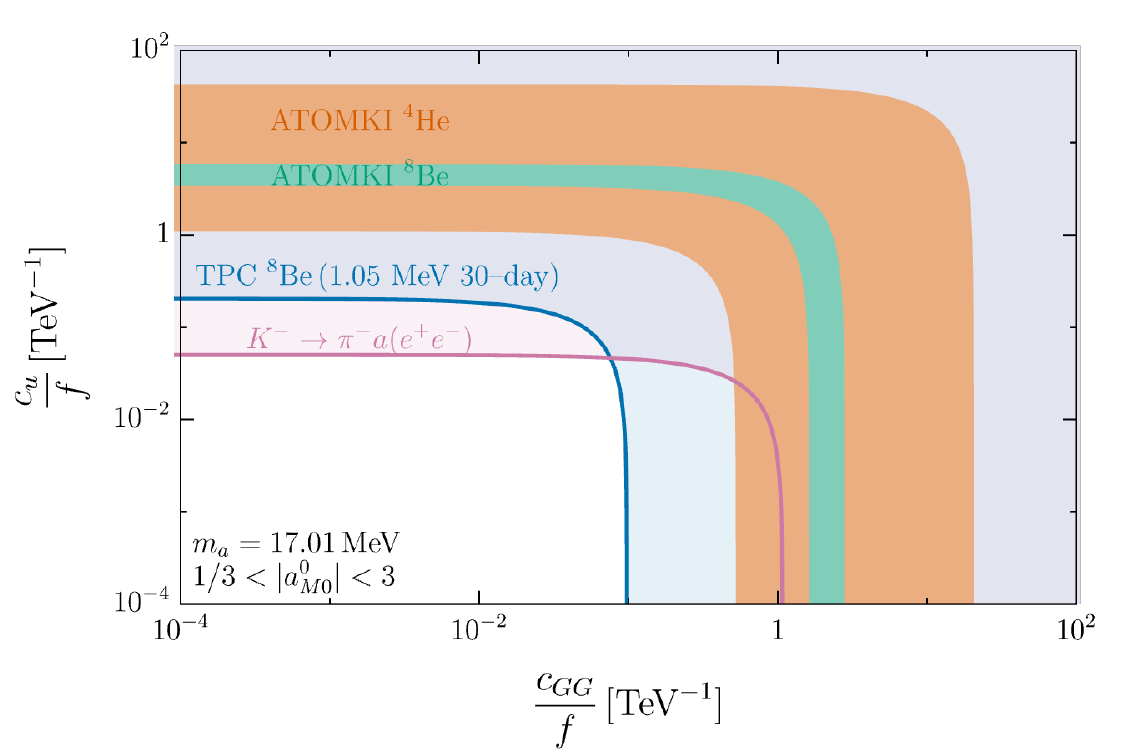}
  \caption{Constraints on ALP couplings in the $c_{GG}-c_u$ plane, for an ALP mass of $m_a=17.01$~MeV, with $c_e/f$ or $c_L/f \gtrsim 11.3\,\text{TeV}^{-1}$ and all other Wilson coefficients set to zero. The green (orange) band shows the region consistent with the ATOMKI observation of the $^8\text{Be}(18.15)$ ($^4\text{He}(21.01)$) transition. The blue shaded region would be excluded by a $30$ day Pelletron run at $E_p=1.05\,$MeV. The pink region is excluded by $K^- \to \pi^- a (e^+ e^-)$ decays \cite{Baker:1987gp}.}
  \label{fig:ALPs2}
\end{figure}

The ALP-nucleon coupling can be related to the Wilson coefficients of \cref{Leff} defined at the UV scale via renormalization group evolution \cite{Bauer:2020jbp,Chala:2020wvs}. Assuming flavour-universal ALP couplings in the UV the isosinglet ALP-nucleon coupling is given by \cite{Bauer}
\begin{align} \label{eq:gaNNNum}
 g_{aNN}^{(0)} = 10^{-4}\, \left[ \frac{1\,\text{TeV}}{f} \right] 
   \times \Big[  &- 4.2 \,c_{GG} + 9.7\times 10^{-4}\,c_{WW} + 9.7\times 10^{-5}\,c_{BB}   \notag\\ 
    &- 2.0\,c_u(\Lambda) - 2.0\,c_d(\Lambda) +4.0\,c_Q(\Lambda)  \notag\\
    &+ 2.9\times 10^{-4}\,c_e(\Lambda) - 1.6\times10^{-3}\,c_L(\Lambda) \Big]^2\,.
\end{align}
Using this relation, we can now show the reach of the TPC on the UV parameter space and compare to independent experimental measurements constraining the same parameters. In \cref{fig:ALPs2}, we show the dominant constraint on the $c_{GG}-c_u$ plane, for $m_a=17$~MeV, assuming that $c_e/f$ or $c_L/f \gtrsim 11.3\,\text{TeV}^{-1}$ and all the other UV Wilson coefficients are zero. We focus in this particular parameter space since it is the only one with a viable region explaining the ATOMKI helium anomaly~\cite{Bauer}.\footnote{Note that such large ALP couplings to electrons are already severely constrained by beam dump experiments. These bounds may, however, be weakened by the inclusion of ALP-quark and -gluon couplings as required here.} Other coupling combinations and ALP couplings to electrons smaller than $c_{e}/f = 11.2\,\text{TeV}^{-1}$ or $c_{L}/f = 11.3\,\text{TeV}^{-1}$ \cite{Bauer} are excluded predominantly by $K_L \to \pi^0 X$ decays~\cite{CortinaGil:2021nts}, with $X$ decaying invisibly or escaping the NA62 detector. In \cref{fig:ALPs2}, the green and orange bands show the $3\sigma$ regions consistent with the ATOMKI measurements of the $^8\text{Be}(18.15)$ and $^4\text{He}(21.01)$ transitions, respectively. The pink region is excluded by $K^- \to \pi^- a (e^+ e^-)$ decays \cite{Baker:1987gp}. For better readability we omit the weaker constraints from $K_L \to \pi^0 X$~\cite{CortinaGil:2021nts} and $K^- \to \pi^- \gamma \gamma$ decays \cite{E949:2005qiy}.
The blue region would be excluded by a $30$ day TPC run at $E_p = 1.05\,$MeV. Note that the projected TPC limit based on an observation of the Beryllium transition alone is capable of excluding the ALP explanation of \emph{both} the ATOMKI Beryllium and Helium observations and also provides the leading constraint in part of the plane.

\section{Future Research Plans}
\begin{table}
\centering
\begin{tabular}{ |c|c|c| } 
 \hline
 Reaction & Q-value & Mass Range for search \\ \hline 
  $p+ ^{7}$Li$\to ^{8}$Be$ \:+ (e^{+} e^{-})$ & 17.25 & 10 - 20 MeV \\ 
\hline
 $p+ ^{3}$H$\to ^{4}$He$ \:+ (e^{+} e^{-})$ & 19.81 & 17 - 22 MeV \\ 
 $p+ ^{11}$B$\to ^{12}$C$ \:+ (e^{+} e^{-})$ & 15.96 & 9 - 19 MeV \\ 
  $p+ ^{27}$Al$\to ^{28}$Si$ \:+  (e^{+} e^{-})$ & 11.57 & 9 - 15 MeV \\     
  $p+ ^{25}$Mg$\to ^{26}$Al$ \:+ (e^{+} e^{-})$ & 6.31 & 5 - 10 MeV \\ 
    $p+ ^{12}$C$\to ^{13}$N$ \:+ (e^{+} e^{-})$ & 1.94 & 3 - 5.5 MeV \\ 
 \hline
\end{tabular}

\caption{Indicative program for future searches for new weakly coupled bosons.  While an inital run would focus on $p\,+\,{}^7\text{Li}$ collisions, potential targets at subsequent runs are also shown. 
`Q-value' is the energy difference between the proton plus target nucleus and the final state nucleus.  
}
\label{table:program}
\end{table}

The nuclear reaction approach has an inherent advantage over particle physics experiments in that it is possible to tune the end-point of the IPC background via choice of nuclear reactions.  For example, The NA48/2 result is limited at low mass by the $\pi^{0} \to \gamma e^{+} e^{-}$ background.  Lower mass searches would be better made with nuclear reactions which populate lower excitation energies in the final state, eg. $p+ ^{13}$C$ \,\to\, ^{13}$N$ + (e^{+} e^{-})$ and $p+ ^{27}$Al$ \,\to\, ^{28}$Si$ + (e^{+} e^{-})$ to limit the IPC background. The IPC background implies that searches for lower mass bosons are best done with reactions with smaller Q-values which have a corresponding lower endpoint for the IPC. 
Higher mass regions can be probed using the $p+^{3}$H$\,\to\, ^{4}$He$+e^{+}e^{-}$ reaction and higher beam energies such as those provided by the ANU 14 UD tandem accelerator.

\begin{table}
    \centering
    \begin{tabular}{|c|cccc|}
    \hline
         $J_\ast^{P_\ast}$ & Scalar $X$ & Pseudoscalar $X$ & Vector $X$ & Axial Vector $X$ \\
         \hline
         $0^-$ & $-$ & $\checkmark$ & $-$ & $\checkmark$\\ 
         $0^+$ & $\checkmark$ & $-$ & $\checkmark$ & $-$\\ 
         $1^-$ & $\checkmark$ & $-$ & $\checkmark$ & $\checkmark$\\
         $1^+$ & $-$ & $\checkmark$ & $\checkmark$ & $\checkmark$\\ 
         \hline
         
    \end{tabular}
    \caption{Allowed spin-parities of the $X$ particle in different transitions $N_\ast \to N$, for $J^P = 0^+$ ground states (such as $^8$Be and $^4$He).}
    \label{tab:allowed-resonances}
\end{table}

An indicative program of reactions covering the 5 -- 22 MeV mass range is shown in \cref{table:program}. The exact choice of energies will be chosen to populate well-known resonances in the final states, particularly for $p+ ^{12}$C$\,\to\, ^{13}$N$ + (e^{+} e^{-})$. In selecting the reaction and excition energies we will be cognascent of the selection rules for differing types of new physics bosons and the spins and parities of the target resonance and final state ground state such as shown in table \ref{tab:allowed-resonances}

Should we find candidate bosons, we can determine their spin and parity via the angular distributions of the $(e^{+}e^{-})$ combined with knowledge of the spin and parity of the initial and final nuclear states.

\section{Conclusions}
%
We propose to build a state-of-the-art time projection chamber integrated with the University of Melbourne Pelletron to perform a series of experiments to resolve the conjecture of the X17 particle and to search for new physics originating from bosons in the 5 - 21 MeV mass range that feebly couple to nuclei. The TPC will be constructed and installed on the University of Melbourne Pelletron in one year following funding. We expect to provide a definitive result on the existence of the X17 one year after completion of construction. Assuming that we do not find the X17, we will begin the longer-term search for feebly interacting bosons using proton-induced nuclear reactions with a variety of tagets and proton energies.

\section*{Acknowledgements}

The authors would like to thank 
Tim Grey (ANU) for the calculation of the energy and angular correlations of the electron-positron
pairs from the normal electromagnetic decay (IPC) of the 18.15 MeV M1 transition in $^{8}$Be.

We thank Nicholas Jackson (Melbourne), who studied the design of the TPC for his M.Sc.~thesis. The COMSOL, GEANT4, GENFIT, RAVE and GARFIELD calculations resulted from his work.

We thank Eraldo Oliveri, Rui de Oliveria, Bertrand Mehl, and Hans Muller, all from CERN, for their advice on the construction and signal readout of the TPC.

We thank Kimika Uehara (Melbourne) who constructed and tested the prototype TPC target chamber as a summer research project in 2022.

This research was partially supported by the Australian Government through the Australian Research Council Centre of Excellence for Dark Matter Particle Physics (CDM, CE200100008).



\begin{thebibliography}{12}


\bibitem{Lanfranchi:2020crw}
G.~Lanfranchi, M.~Pospelov and P.~Schuster,
\href{https://www.annualreviews.org/doi/10.1146/annurev-nucl-102419-055056}{Ann. Rev. Nucl. Part. Sci. \textbf{71} (2021), 279-313}

\bibitem{Agrawal:2021dbo}
P.~Agrawal, M.~Bauer, J.~Beacham \textit{et al.}
\href{https://link.springer.com/article/10.1140/epjc/s10052-021-09703-7}{Eur. Phys. J. C \textbf{81} (2021) no.11, 1015}

\bibitem{2016Kr_PRL}
    A.J.~Krasznahorkay, \emph{et al.}, 
     \href{https://link.aps.org/doi/10.1103/PhysRevLett.116.042501}
     {Phys. Rev. Lett. {\bf 116}, 042501 (2016)}

\bibitem{2021Kr02_PRC}
A.J.~Krasznahorkay, \emph{et al.}, 
\href{https://doi.org/10.1103/PhysRevC.104.044003} 
{Phys. Rev. {\bf C104}, 044003 (2021)}

\bibitem{2023Kr03_PRC}
A.J.~Krasznahorkay, \emph{et al.}, 
\href{https://arxiv.org/abs/2209.10795v2}{Phys. Rev {\bf C }(In press), 2023} 


\bibitem{2020Banerjee_PRD}
  D.~Banerjee, \emph{et al.},
  \href{https://link.aps.org/doi/10.1103/PhysRevD.101.071101}
  {Phys. Rev. D {\bf 101}, 071101(R) (2020)}

\bibitem{2021Aleksejevs_arXiv}
    A.~Aleksejevs, {\emph et al.}, 
    \href{https://arxiv.org/abs/2102.01127v1}{arXiv:2102.01127v1}

  \bibitem{2022Viviani_PRC}
  M. Viviani, \emph{et al.},
  \href{https://link.aps.org/doi/10.1103/PhysRevC.105.014001}
  {Phys. Rev. C {\bf105}, 014001 (2022)}
  
\bibitem{2020Feng_PRD}
  J.L.~Feng, T.M.P.~Tait, C.B.~Verhaaren,
  \href{https://doi.org/10.1103/PhysRevD.102.036016}
  {Phys. Rev. D {\bf 102}, 036016 (2020)}

\bibitem{2021Zhang_PLB}
    X.~Zhang and G.~A.~Miller,
    \href{https://doi.org/10.1016/j.physletb.2021.136061}
    {Phys.~Lett. {\bf 813}, 136061 (2021)}


\bibitem{2016Feng_PRL}
     J.L.~Feng,  \emph{et al.}, 
    \href{https://link.aps.org/doi/10.1103/PhysRevLett.117.071803}
    {Phys. Rev. Lett. {\bf 117}, 071803 (2016)}

\bibitem{Boehm:2002yz} 
  C.~Boehm, T.A.~Ensslin and J.~Silk,
  \href{https://doi.org/10.1088/0954-3899/30/3/004}
  {J.\ Phys.\ G {\bf 30}, 279 (2004)}
  
\bibitem{Boehm:2003hm}
  C.~Boehm and P.~Fayet,
  \href{https://doi.org/10.1016/j.nuclphysb.2004.01.015}
  {Nucl.\ Phys.\ B {\bf 683}, 219 (2004)}

\bibitem{Alves}
D.~Alves,
\href{https://doi.org/10.1103/PhysRevD.103.055018}{Phys. Rev. {\bf D103} 055018 (2021)}

\bibitem{EVE}
M.~Tadel, 
\href{https://doi.org/10.1088/1742-6596/219/4/042055}
{J. Phys.: Conf. Ser. 219 042055 (2010)}


\bibitem{root}
R.~Brun and F.~Rademakers, 
\href{https://doi.org/10.1016/S0168-9002(97)00048-X}{Nucl. Inst. \& Meth. in Phys. Res. {\bf A 389} 81-86 (1997)}

\bibitem{micromegas}
Y.~Giomataris, Ph.~Rebourgearda, J.P.~Robert and G.~Charpak,
\href{https://doi.org/10.1016/0168-9002(96)00175-1}
{Nucl.~Instr.~and~Meth.~Phys.Res.,{\bf 376}, 29, (1996)}

\bibitem{comsol}
E.~Dickinson, H.~Ekström, E.~Fontes,
\href{https://doi.org/10.1016/j.elecom.2013.12.020}{Electrochemistry Communications {\bf 40}, 71-74 (2014)}
 
\bibitem{garfield}
R.~Veenhof,
\href{https://doi.org/10.1016/S0168-9002(98)00851-1}{Nucl. Inst. \& Meth. in Phys. Res. {\bf A 419} 726-730 (1998)}

\bibitem{vmm}
 T. Alexopoulos, \emph{et al.},
 \href{http://doi.org/10.1109/NSS/MIC42677.2020.9508044}{2020 IEEE Nuclear Science Symposium and Medical Imaging Conference, 1-3,(2020)}
   P.~Gkountoumis, 
  \href{https://doi.org/10.1088/1748-0221/12/01/C01088}
  {JINST {\bf 12} C01088 (2017)}


\bibitem{gdml}
R. Chytracek, \emph{et al.}, 
\href{http://doi.org/10.1109/TNS.2006.881062/}{IEEE Trans. on Nucl. Sci. {\bf 53} 2892 - 2896 (2006)}

\bibitem{GENFIT}
  Belle~II~Tracking~Group, V.~Bertacchi, T.~Bilka, \emph{et al.},
  \href{https://doi.org/10.1016/j.cpc.2020.107610}
  {Comp.~Phys.~Comm. {\bf 259}, 107610 (2021)}

\bibitem{rave}
W. Waltenberger, W. Mitaroff and F. Moser,
\href{https://doi.org/10.1016/j.nima.2007.08.048}{Nucl. Inst. \& Meth. in Phys. Res. {\bf A 581} 549-552, (2007)}

\bibitem{RD51}  
    S.~Martoiu, H.~Muller and J.~Toledo, 
    \href{https://doi.org/10.1109/NSSMIC.2011.6154414}
    {2011 IEEE Nucl. Sci. Symp. Conf. Rec., Valencia, 2011, pp. 203}

\bibitem{Zahnow}
D.~Zahnow, C.~Angulo, C.~Rolfs \emph{et al.},
\href{https://doi.org/10.1007/BF01289534}{Z. Phys. {\bf A 351},229-236 (1995)}


\bibitem{fisher}
G. A.~Fisher, P.~Paul, F.~Riess, and S.S.~Hanna,
\href{https://doi.org/10.1103/PhysRevC.14.28}{Phys. Rev. {\bf C14} 22-36 (1976)}

\bibitem{2017Feng_PRD}
   J.L.~Feng, \emph{et al.}, 
   \href{https://link.aps.org/doi/10.1103/PhysRevD.95.035017}
   {Phys.~Rev.~D {\bf 95}, 035017 (2017)}

\bibitem{Holdom:1985ag}
B.~Holdom,
\href{https://www.sciencedirect.com/science/article/pii/0370269386913778?via%3Dihub}{Phys. Lett. B \textbf{166} (1986), 196-198}

\bibitem{Fabbrichesi:2020wbt}
M.~Fabbrichesi, E.~Gabrielli and G.~Lanfranchi,
\href{https://link.springer.com/book/10.1007/978-3-030-62519-1}{ Springer Briefs in Physics}

\bibitem{NA48}
    J.R.~Batley, \emph{et al.},
     \href{http://dx.doi.org/10.1016/j.physletb.2015.04.068}
     {Phys. Lett. {\bf B 746}, 178-185 (2015)}


\bibitem{E141}
 S.~ Andreas, C.~ Niebuhr, A.~ Ringwald, \href{https://doi.org/10.1103/PhysRevD.86.095019}{Phys. Rev. {\bf D 86} 095019 (2012)}

\bibitem{electron_g-2}
M. Pospelov,
\href{https://doi.org/10.1103/PhysRevD.80.095002}{Phys. Rev. D 80 (2009) 095002}

\bibitem{Ellwanger:2016wfe}
U. Ellwanger, S. Moretti, 
\href{https://doi.org/10.1007/JHEP11(2016)039}{J. High Energ. Phys. 2016, 39 (2016)} 

\bibitem{Donnelly:1978ty}
T.W.~Donnelly \emph{et al.},
\href{https://doi.org/10.1103/PhysRevD.18.1607}{Phys. Rev. D, {\bf 18}, 1607 (1978)}

\bibitem{Barroso:1981bp}
A.~Barroso and N.~Mukhopadhyay,
\href{https://doi.org/10.1103/PhysRevC.24.2382}{Phys. Rev. C, \bf{24}, 2382 (1981)}


\bibitem{Bauer:2020jbp}
M.~Bauer, M.~Neubert, S.~Renner \emph{et al.}, 
\href{https://link.springer.com/article/10.1007/JHEP04(2021)063}{JHEP \textbf{04} (2021), 063}

\bibitem{Chala:2020wvs}
M.~Chala, G.~Guedes, M.~Ramos \emph{et al.},
\href{https://link.springer.com/article/10.1140/epjc/s10052-021-08968-2}{Eur. Phys. J. C \textbf{81} (2021) no.2, 181}

\bibitem{Bauer}
M.~Bauer, M.~Neubert, S.~Renner \emph{et al.}, 
\href{https://doi.org/10.1007/JHEP09(2022)056}{J. High Energ. Phys. 2022, 56 (2022)}

\bibitem{Baker:1987gp}
N.J.~Baker, \emph{et al.},
\href{https://doi.org/10.1103/PhysRevLett.59.2832}{Phys. Rev. Lett. {\bf 59}, 2832 (1987)}

\bibitem{CortinaGil:2021nts}
E.~Cortina Gil \emph{et al.},
\href{https://doi.org/10.48550/arXiv.2103.15389}{J. High Energ. Phys. 06, 093 (2022)}


\bibitem{E949:2005qiy}
A.~V.~Artamonov \textit{et al.}, [E949],
\href{https://www.sciencedirect.com/science/article/abs/pii/S0370269305010427?via%3Dihub}{Phys. Lett. B \textbf{623} (2005), 192-199}

\end{thebibliography}
\end{document}